\documentclass{aa}  

\usepackage{graphicx}
\usepackage{txfonts}
\usepackage{multirow}
\usepackage{booktabs}
\usepackage{makecell}
\usepackage{xcolor}         
\usepackage{amsmath}        
\usepackage{tikz}           
\usetikzlibrary{positioning, shapes, arrows.meta, decorations.pathreplacing}

\def\astrolink{\texttt{AstroLink}}
\makeatletter
\newcommand*{\defeq}{\mathrel{\rlap{%
                     \raisebox{0.3ex}{$\m@th\cdot$}}%
                     \raisebox{-0.3ex}{$\m@th\cdot$}}%
                     =}
\makeatother
\usepackage{hyperref}
\begin{document}

   \title{The hierarchical structure of the Milky Way}

   \subtitle{I. A model-agnostic clustering of 1 billion stars in 5D}

   \author{William H. Oliver
          \inst{1,2}
          \and 
          Tobias Buck
          \inst{1,2}
          }

   \institute{Interdisziplin\"{a}res Zentrum f\"{u}r Wissenschaftliches Rechnen, Universit\"{a}t Heidelberg, Im Neuenheimer Feld 205, D-69120 Heidelberg, Germany\\
             \email{william.hardie.oliver@gmail.com, tobias.buck@iwr.uni-heidelberg.de}
             \and
             Universität Heidelberg, Zentrum für Astronomie, Institut für Theoretische Astrophysik, Albert-Ueberle-Straße 2, D-69120 Heidelberg, Germany
             }

   \date{}

 
    \abstract
    {The Milky Way exhibits a hierarchical structure shaped by mergers, accretion, and internal dynamics. Gaia Data Release 3 (DR3) enables a global view of this structure, but requires scalable, model-agnostic methods to extract multi-scale clustering patterns.}
    {We aim to characterise the hierarchical clustering structure of the Milky Way in a model-agnostic way and to assess how the resulting structures relate to existing catalogues.}
    {We apply the hierarchical density-based clustering algorithm \astrolink\ to a locally magnitude-complete subsample of over one billion Gaia DR3 stars with five-dimensional astrometric information. The data are corrected for selection effects and transformed using a distance contraction and feature scaling. We compare the resulting cluster hierarchy to catalogues of open clusters, globular clusters, dwarf galaxies, and stellar streams using custom availability-aware metrics and a bi-directional best-match framework.}
    {At a significance threshold of $S = 3.8\sigma$, \astrolink\ identifies 7410 statistically significant clusters spanning a wide range of morphologies. Compact systems are often recovered as reciprocal matches with many also exhibiting internal substructure, while extended structures are typically fragmented or partially recovered. A substantial fraction of clusters have no counterpart in the comparison catalogues, these may be new structures but additional physical checks are needed to verify this.}
    {A model-agnostic, hierarchical clustering approach can recover the multi-scale structure of the Milky Way at Gaia DR3 scale and provides a unified framework for comparing heterogeneous catalogues and identifying candidate structures.}

    \keywords{Galaxy: structure -- 
              Galaxy: stellar content -- 
              Galaxy: kinematics and dynamics -- 
              methods: data analysis -- 
              methods: statistical
              }

   \maketitle
%

\section{Introduction}
The Milky Way (MW) is a dynamically evolving, hierarchical system whose present-day structure encodes its formation history. Over cosmic time, it has grown through the accretion and disruption of smaller systems \citep{White1978, Searle1978, Kauffmann1993, Bullock2005}, while internal processes have shaped its disc, halo, and bulge \citep{Sellwood2002, Kormendy2004, Minchev2014, Obreja2019, Buck2018, Buck2019b}. The resulting stellar distribution exhibits structure across a wide range of spatial and kinematic scales, from compact bound systems such as open and globular clusters to extended tidal streams, dwarf galaxies, and diffuse components of the galactic disc and halo \citep{Freeman2002, Helmi2020}. Characterising this hierarchy of structure is central to galactic archaeology, as it provides direct insight into past merger events, dynamical evolution, and the underlying cosmological model.

The advent of large-scale stellar surveys -- most notably; SDSS \citep{York2000, Kollmeier2026}, DES \citep{DES2016, Abbott2018}, Gaia \citep{Prusti2016, GaiaDR3}, GALAH \citep{deSilva2015, Buder2025}, APOGEE \citep{Majewski2017}, and Pan-STARRS \citep{Chambers2016} -- has transformed our ability to study this structure \citep{BlandHawthorn2016, Helmi2020}. These surveys have provided precise positions, proper motions, distances, and chemistries for many stars across the sky, enabling a global and homogeneous view of the MW. This wealth of data has led to the discovery and refinement of numerous stellar systems, as well as to the identification of previously unseen substructure in both the disc and halo \citep{PriceWhelan2018, Shipp2018, Belokurov2018, Helmi2018}. At the same time, the scale and complexity of these data sets present significant challenges, i.e. extracting meaningful structure from such a high-dimensional and heterogeneous data set requires methods that are both computationally efficient and robust to variations in density, morphology, and observational uncertainty.

A wide range of approaches have been developed to identify stellar structures in Gaia data. These typically make use of widely used centroid-based, model-based, or density-based clustering algorithms such as K-Means \citep{MacQueen1967}, EM \citep{Dempster1977}, DBSCAN \citep{Ester1996}, and HDBSCAN \citep{Campello2015, McInnes2017}. However there are also a growing number of fit-for-purpose clustering algorithms that use similar algorithmic concepts, e.g. Matched Filter \citep{Rockosi2002, Balbinot2011}, xGC3 \citep{Johnston1996, Mateu2011, Mateu2017}, StreamFinder \citep{Malhan2018a}, StarGo \citep{Yuan2018}, HSS \citep{Pearson2022}, Via Machinae \citep{Shih2022, Shih2024}, IOM \citep{Lovdal2022}, SigMA \citep{Ratzenboeck2023}, SKYCURTAINS \citep{Sengupta2025}, StarStream \citep{Chen2025}, etc. These methods have been applied in the context of; open cluster searches \citep{CantatGaudin2018, Hunt2021, Hunt2024}, (chemo-)kinematic analyses of globular clusters and dwarf galaxies \citep{Willman2005, Belokurov2006, Koposov2008, Minniti2017, Gran2019, Torrealba2019}, and track-based or phase-space searches for stellar streams \citep{Johnston1996, Mateu2011, Mateu2017, Ibata2019, Ibata2021, Malhan2022}.

These algorithms and their applications are typically tailored to specific object classes and often rely on assumptions about morphology, scale, or dynamical coherence. Consequently, Galactic structures are often identified and analysed in isolation with respect to a specific methodology, selection criteria, and membership definition. As a result, it is difficult to construct a unified and internally consistent view of the MW’s structure that reflects both its intrinsic multi-scale nature and its broad taxonomy. A key limitation of existing approaches is therefore the lack of a model-agnostic framework that can (1) recover diverse structure types across all relevant scales simultaneously, and (2) crossmatch against existing catalogues without relying on metrics that are dependent on catalogue completeness or model-based assumptions.

In this work, we address this gap by applying the hierarchical, model-agnostic clustering algorithm \astrolink\ \citep{Oliver2024a} to a large Gaia DR3 subsample containing over one billion stars with five-dimensional astrometric information. By identifying overdensities of arbitrary shape and organising them into a multi-scale hierarchy, \astrolink\ is well suited to the heterogeneous structure of the MW. In addition to recovering statistically significant structures across a wide range of morphologies and scales, we develop a model-agnostic framework for comparing the resulting hierarchy to existing catalogues through availability-aware metrics and bi-directional best-match relations. This enables a consistent structural taxonomy describing reciprocal matches, fragmentation, merging, and other associations between \astrolink\ clusters and known stellar systems.

The structure of this paper is as follows. In Sec. \ref{sec:astrolink} we describe the \astrolink\ algorithm and the methodological extensions used in this work. Sec. \ref{sec:data} outlines the construction of the Gaia DR3 subsample and the transformations applied to the data. In Sec. \ref{sec:results_first_look} we present a first overview of the clustering results. Sec. \ref{sec:catalogue_crossmatching} is devoted to the comparison with external catalogues and the analysis of the resulting structural taxonomy. Finally, in Sec. \ref{sec:conclusions} we summarise our findings and discuss prospects for future work.

\section{Methods: Extracting clusters from the data} \label{sec:astrolink}

To extract structure from our Gaia DR3 subsample (see Sec.~\ref{sec:data} for data set details), we use the hierarchical density-based astrophysical clustering algorithm, \astrolink.

\subsection{The AstroLink algorithm}
\astrolink\ \citep{Oliver2024a} is an unsupervised astrophysical clustering algorithm that is able to extract a hierarchy of arbitrarily-shaped clusters from any point-based data set. The clusters themselves are collections of data points, each defining a feature space overdensity that is statistically distinct from the stochastic clumping within the data. \astrolink\ computes these clusters and the hierarchy they belong to in five steps;
\begin{enumerate}
    \item If the hyperparameter \texttt{adaptive} $=1$ (default), \astrolink\ first rescales the input data so that each dimension has unit variance. This addresses the effect of differing units in the feature space dimensions.
    \item \astrolink\ then calculates the local-density of each data point by applying a multivariate Epanechnikov kernel \citep{Epanechnikov1969} and a balloon estimator \citep{Sain2002} to its $k_\mathrm{den}$-neighbourhood ($k_\mathrm{den}=20$, default) -- the logarithm of this estimate is taken, after which all values are normalised, so that $\log\hat\rho \in [0, 1]$.
    \item Now a local-density-weighted $k_{\mathrm{link}}$-nearest-neighbour graph ($k_\mathrm{link}$ is data-driven by default and is selected to ensure graph-connectedness in $10^5$ points) is built and then traversed in descending order of edge weight. The connected components of data points that form during the latter process define a hierarchy of feature-space overdensities.
    \item Then, a model is fit to the prominence (a measure of \textit{clusteredness}) of these connected components and is used to identify the $\geq S\sigma$-outlier overdensities ($S$ is data-driven by default and is selected so that noisy groups are not expected to appear as clusters) from the noisy local-density fluctuations inherent within the data.
    \item Lastly, if the hyperparameter \texttt{h\_style} $=1$ (default), a correction is applied to the hierarchy clusters by incorporating some additional outlier overdensities -- producing the final hierarchy of clusters.
\end{enumerate}

For more algorithmic details, we refer the reader to the original science paper \citep{Oliver2024a}, as well as to the \astrolink\ GitHub (\url{github.com/william-h-oliver/astrolink}) and ReadTheDocs (\url{astrolink.readthedocs.io}) pages.

Compared to its predecessors, \texttt{CluSTAR-ND} \citep{Oliver2022} and \texttt{Halo-OPTICS} \citep{Oliver2020}, as well to traditional halo-finders like AHF \citep{Gill2004, Knollmann2009}, \astrolink\ is shown to perform better at finding infalling and disrupted satellite remnants within simulated galaxies while also taking less run-time \citep{Oliver2024a, Oliver2024b, Ekanayaka2025, Babu2026}. It has also recently been applied to Gaia DR3 to derive membership lists for Milky Way open clusters as part of an updated catalogue of cluster ages and parameters \citep{Kallimanis2026}.

Typically, \astrolink\ does not require the user to make any hyperparameter choices as the performance of the entirely data-driven version of this process is near-optimal in most cases. However in the case of applying \astrolink\ to large and non-idealised observational data such as the Gaia DR3 subsample discussed in this paper, there are specific considerations to make -- hence some small functionality changes have been implemented within \astrolink\ that differ from previous versions.


\subsection{Correcting for observational biases} \label{subsec:selectionfunctionreweighting}

\astrolink\ gives robust and meaningful results whenever its input data can be interpreted as a random sample of the true underlying distribution on some domain of informative features. For idealised and reliable simulated data, the user needs only to consider whether the feature space is an informative one. However, as is discussed in Sec.~\ref{subsec:selectionfunction}, an observational data set is generally a biased sample of the true distribution -- which in our case takes the form of $S(\textbf{x})P_\mathrm{true}(\textbf{x})$ with selection function $S(\textbf{x})$. Therefore an estimate of the true distribution can be recovered by weighting the input data $\textbf{x}$ with a factor of $S(\textbf{x})^{-1}$.

\begin{figure}
    \centering
    \vspace{-0.25cm}
    \includegraphics[trim={5mm 3.25mm 3.5mm 3.5mm}, clip, width=\columnwidth]{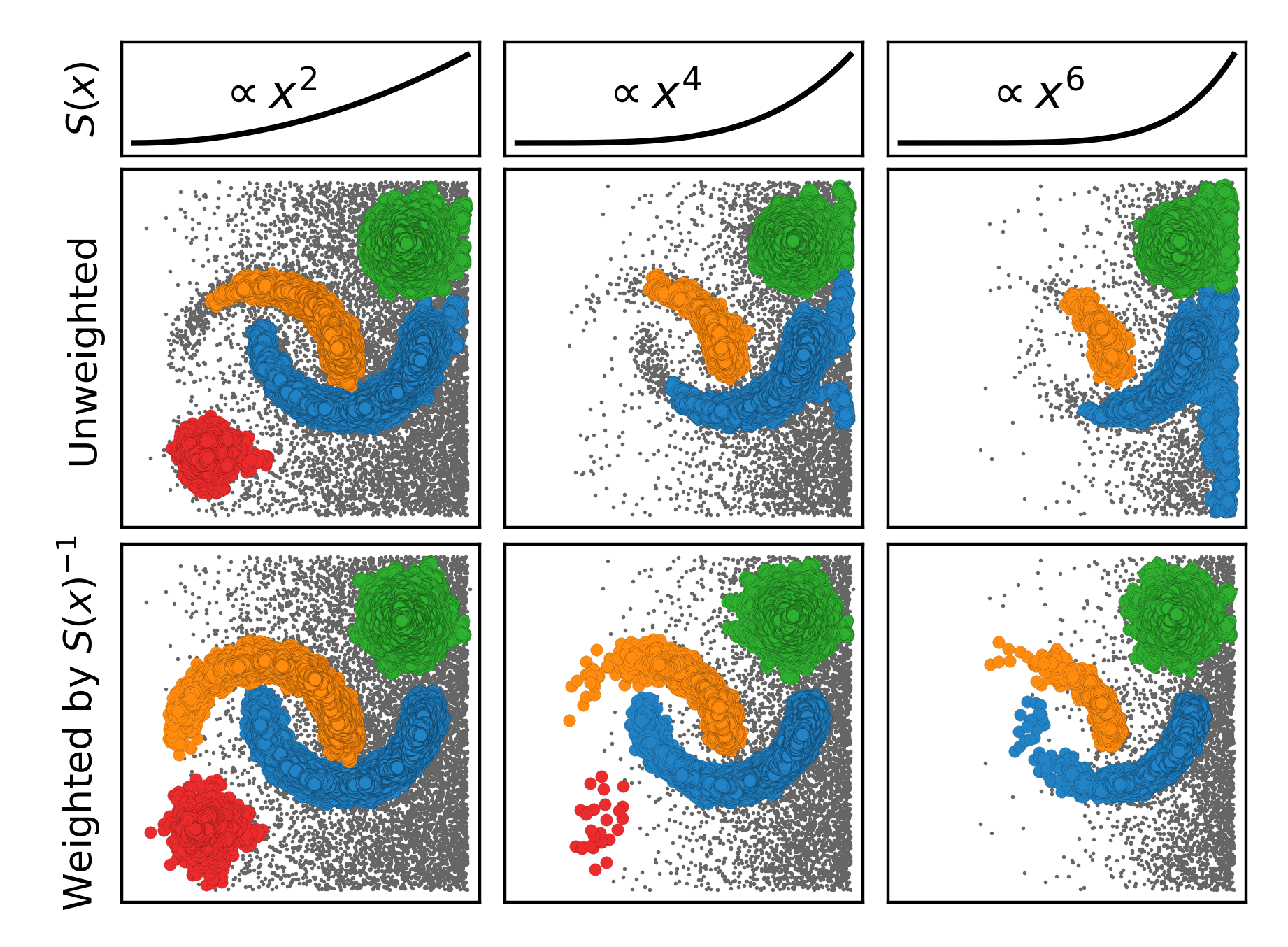}
    \vspace{-0.25cm}
    \caption{The effect of appropriately weighting input data on the \astrolink\ output. Here various subsamples of a 2D toy data set, consisting of two half-moons and two Gaussian blobs within a uniform background distribution, are taken using different selections functions (top row). Then \astrolink\ is applied to both unweighted (middle row) and weighted (bottom row) versions of these subsamples. While the unweighted clusterings tend to exhibit both poor recovery and purity simultaneously, the weighted clusterings are mostly unaffected (except in the limit of very small or zero counts) by the apparent density gradient imposed by the selection functions.}
    \label{fig:effect_of_selection_function}
    \vspace{-0.25cm}
\end{figure}

As such, the user can now give a \texttt{weights} array alongside the input data which can be used to adjust for observational biases (or for differing particle masses in the case of a simulated data set). If these weights are provided, then the local-density of each point is computed using 
\begin{align} \label{eq:densityestimator}
    &\hat\rho_i \propto \frac{1}{h_i^d} \sum_{j \in N_{k_\mathrm{den}}} w_j K\left(\frac{s(x_i, x_j)}{h_i}\right),\ \mathrm{where} \notag\\
    &h_i = \mathrm{max}\{s(x_i, x_j) \mid j \in N_{k_\mathrm{den}}\}\ \mathrm{and}\ K(u) \propto 1 - u^2.
\end{align}
Here $w_j$ is the weight of point $j$, $d$ is the dimensionality of the feature space, $N_{k_\mathrm{den}}$ is the set of point $i$'s $k_\mathrm{den}$-nearest-neighbours, and $s(x_i, x_j)$ is the Euclidean distance between points $i$ and $j$ which, if {\tt adaptive} $= 1$, corresponds to the Mahalanobis distance on the input data and is otherwise the Euclidean distance. Here $K(u)$ is the Epanechnikov kernel \citep{Epanechnikov1969}, which is optimal among kernels that are themselves valid probability density functions in the sense that it minimizes the mean integrated squared error (MISE). Note that $K(u) \defeq 0$ for $u > 1$. If weights are not provided then each $w_j$ are treated as unity and Eq.~\ref{eq:densityestimator} becomes the same as Eq. 1 in \citet{Oliver2024a}. An example of the effect that appropriately weighting the input data can have on the clustering power of \astrolink\ is shown in Fig.~\ref{fig:effect_of_selection_function}.

\subsection{Accounting for data embedded on a sub-manifold} \label{subsec:intrinsic_dimensionality}

Another recent change to \astrolink\ is to allow the user to account for data that exists on a sub-manifold within the input data space. For example, if the input data is $n$-dimensional but the distribution of data points lives on an $m$-dimensional subspace ($m < n$), then distances between nearest neighbours will be distributed according to the underlying geometry of this subspace -- assuming that the curvature of the surface within each point's $k_\mathrm{den}$-neighbourhood is sufficiently small. If \astrolink\ is applied to such a data set and the intrinsic dimensionality of the input data is assumed to be $n$, then Eq.~\ref{eq:densityestimator} will result in $\hat\rho_i$ values that are incorrectly multiplied by a factor of $h_i^{m - n}$. So to broaden \astrolink's applicability in these cases, the user can now specify the intrinsic dimensionality using the \texttt{d\_intrinsic} hyperparameter -- which then sets the value of $d$ in Eq.~\ref{eq:densityestimator} accordingly.

\section{Data: An augmented Gaia DR3 subsample} \label{sec:data}
To reveal as much of the MW's structure as possible with \astrolink, we use the positions-on-the-sky ($\alpha$, $\delta$) and proper motions ($\mu_\alpha$, $\mu_\delta$) of stars in the all-sky Gaia DR3 catalogue (GDR3) \citep{Prusti2016, GaiaDR3} as well as the improved photo-geometric distances from \citet{BailerJones2021}. There are $\sim1.35$ billion stars with such solutions, however many of these are considered \textit{unreliable}, for which the standard errors are not expected to encapsulate the true values. In addition to this and in general, an observational data set is only a subset of the true distribution, where the probability of a star being included in the subset is dependent upon a range of properties. If not accounted for, both of these characteristics will adversely affect the determination of the clustering structure within the data. The following subsections give details of how we remove and/or adjust for them to ultimately provide a robust prediction of the MW's clustering structure.

\subsection{Catalogue cleaning} \label{subsec:cataloguecleaning}
We first clean the GDR3 catalogue of stars to ensure that the resulting subsample we use has reliable astrometric solutions. Typically a Gaia $G$-band magnitude cut is used and is commonly combined with additional cuts on other GDR3 quantities such as the $G$-band extinction and the \texttt{ruwe} quality indicator. However, the reliability of a star's astrometric solution is not only related to its $G$-band magnitude but also to the density of sources surrounding it on the sky -- and not independently so. In essence, a star of a given $G$-band magnitude is more/less likely to be well-resolved and tracked over time by the Gaia satellite in less/more crowded regions -- leading to a trend in the quality of the astrometric solution as a function of on-sky position.

This effect has been well-characterised by the empirical survey selection function constructed by \citet{CantatGaudin2023} -- for which many related functionalities are implemented within the \texttt{GaiaUnlimited} python package. This selection function, which we will denote as $S_\mathrm{Gaia}(\alpha, \delta, G)$, provides an estimate of the probability that a star with a given position-on-the-sky and $G$-band magnitude existing in the DECaPS survey \citep{Schlafly2018, Saydjari2023} would appear in the GDR3 catalogue. 
Since the DECaPS survey is a much deeper survey than GDR3 and is considered complete at all Gaia $G$-band magnitudes, $S_\mathrm{Gaia}(\alpha, \delta, G)$ can be taken as the fraction of stars in the MW with attributes $(\alpha, \delta, G)$ that appear in the GDR3 catalogue.

In practice, the median $G$-band magnitude of stars with an \texttt{astrometric\_matched\_transits} (AMT) value of $\leq 10$ ($M_{10}$) is first defined at the position of $(\alpha, \delta)$, and then using this, a fitted generalised sigmoid model is used to provide the probability for a given $G$-band magnitude. The \texttt{GaiaUnlimited} implementations of $S_\mathrm{Gaia}(\alpha, \delta, G)$ all use a HEALPix-based binning scheme to calculate $M_{10}$, where the stars contained within patches on the sky containing the provided $(\alpha, \delta)$ coordinates are used as the set from which the median is taken. This, however, leads to sharp local discontinuities in the $M_{10}$ map as a function of $(\alpha, \delta)$ and hence a discontinuous selection function. So, in order to acquire a smoother $M_{10}$ map, we calculate it uniquely for each star ($i$) in the catalogue using a $k$-nearest-neighbour scheme.

Specifically, and given a position of $(\alpha, \delta)$, we find the $128$-nearest-neighbours on the sky from a reduced set of stars that all have a valid $G$-band magnitude and have $\mathrm{AMT} \leq 10$, i.e. $N_{128}(\alpha, \delta\ |\ \mathrm{valid}\ G \land \mathrm{AMT} \leq 10)$. Using this, we can then estimate $S_\mathrm{Gaia}(\alpha, \delta, G)$ for all stars with valid $G$-band magnitudes by calculating $M_{10}$ and using the \texttt{GaiaUnlimited} \texttt{m10\_to\_completeness} function. Following this we construct our cleaned subsample by requiring that each star have a $S_\mathrm{Gaia}(\alpha, \delta, G)$ value greater than $0.99$, a RUWE value of $<1.2$ (to remove potential binary stars), and a valid 5d astrometric solution. This leaves us with a subsample of 1,032,467,155 stars.

\begin{figure}
    \centering
    \includegraphics[trim={9mm 11mm 9mm 6mm}, clip, width=\columnwidth]{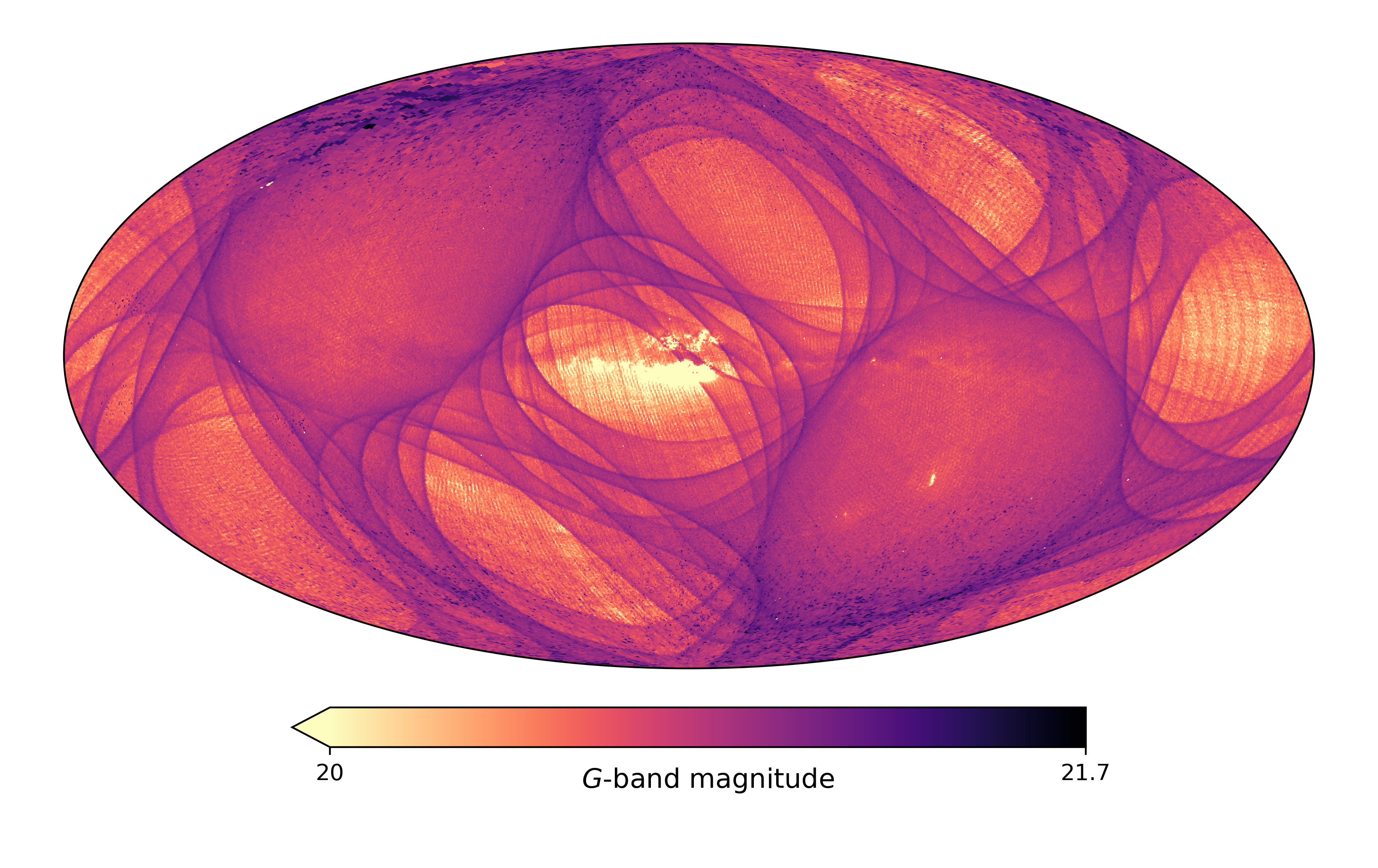}
    \caption{The lower $G$-band magnitude limit as a function of galactic coordinates that we use as a cut to construct our 5d astrometric subsample from the GDR3 catalogue as described in Sec.~\ref{subsec:cataloguecleaning}. We see that the limiting $G$-band magnitude of \textit{reliable} 5d astrometric solutions is a complex combination of both the Gaia satellite scanning law and the crowding of stellar sources from our Galaxy.}
    \label{fig:limitingGband}
\end{figure}

Due to $S_\mathrm{Gaia}(\alpha, \delta, G)$ being a bijective function of $G$, the cut we take here is equivalent to taking a $G$-band cut that varies with the position-on-the-sky as shown in Fig.~\ref{fig:limitingGband}. While other magnitude-complete subsamples are typically constructed by taking all stars with a magnitude less than some \textit{globally} limiting magnitude, our subsample is \textit{locally} magnitude-complete and hence we retain a large number of stars in our subsample. With this high level of completeness everywhere, we now trust that the astrometric solution errors reliably encapsulate the true astrometric solutions of the stars in our subsample.

\subsection{The total selection function of our subsample} \label{subsec:selectionfunction}
Although the astrometric solutions of stars in our subsample may be more reliable as a result of the catalogue cleaning done in Sec.~\ref{subsec:cataloguecleaning}, the distribution of the stars in this feature space is not a random sample of stars taken from the true underlying distribution. Instead our subsample is biased against this true distribution as result of the ability of the Gaia satellite to detect the star given its properties as well as of the cuts we have made to the catalogue. This bias can lead to spurious results when applying any model that does not explicitly account for it -- the effect that such a bias can have on the \astrolink\ clustering results is shown in Fig.~\ref{fig:effect_of_selection_function} for a toy example.

One way in which we can understand and correct for these biases is with the total selection function of our subsample, i.e. the probability that an existing star with given properties is a part of our subsample. Using the ideas introduced by \citet{Rix2021} and \citet{CastroGinard2023}, the total selection function for our subsample can be constructed with 
\begin{equation} \label{eq:totalsf_withG}
    S_\mathrm{total}(\alpha, \delta, G) = S_\mathrm{sub}(\alpha, \delta, G) \cdot S_\mathrm{Gaia}(\alpha, \delta, G)
\end{equation}
where $S_\mathrm{sub}(\alpha, \delta, G)$ is the probability that a star with attributes $(\alpha, \delta, G)$ is selected to be in our subsample from the full GDR3 catalogue and $S_\mathrm{Gaia}(\alpha, \delta, G)$ is the empirical survey selection function as discussed and used in Sec.~\ref{subsec:cataloguecleaning}.

\begin{figure}
    \centering
    \includegraphics[trim={4mm 8.5mm 4mm 2.5mm}, clip, width=\columnwidth]{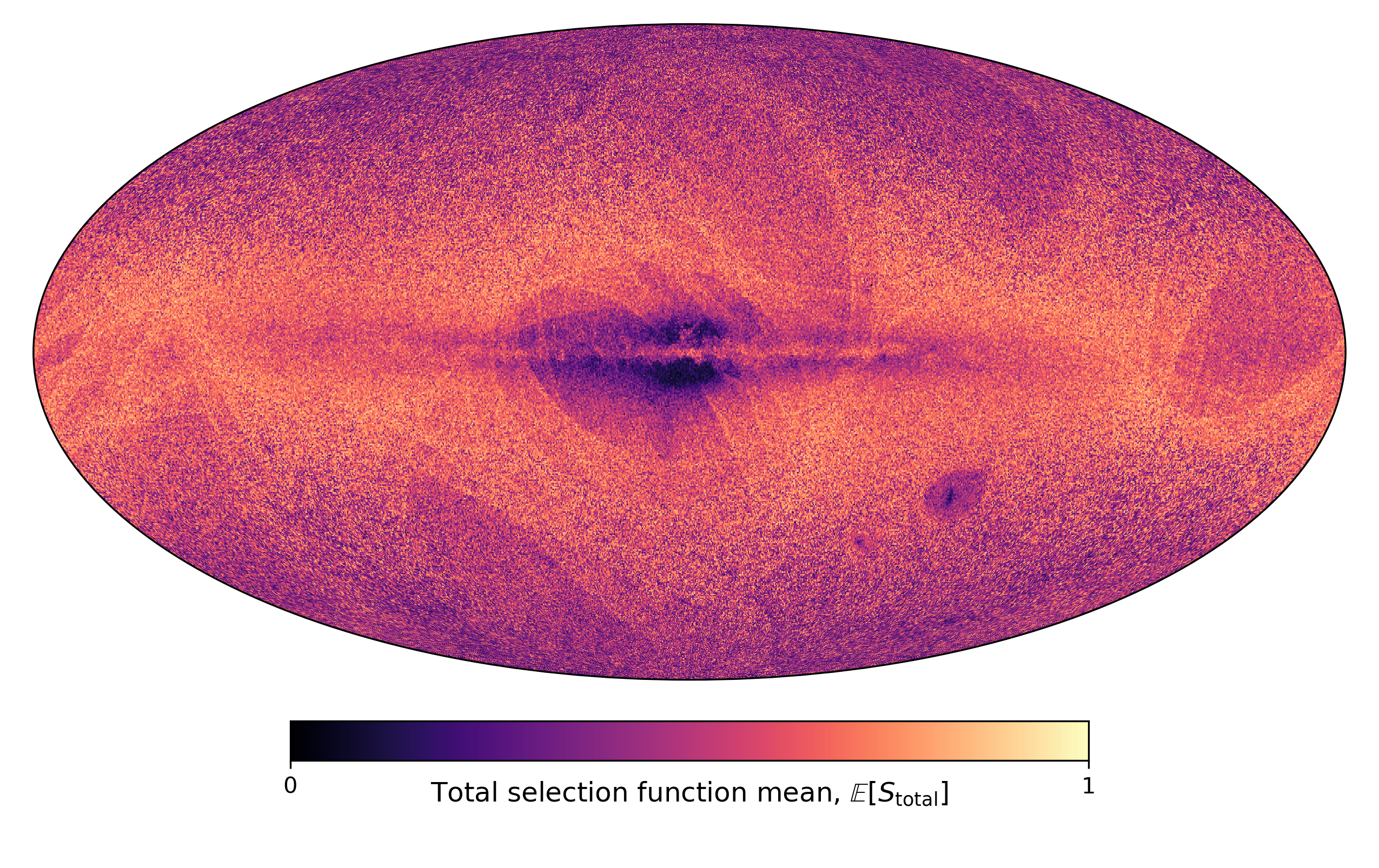}
    \caption{The total selection function of our subsample, $S_\mathrm{total}$, as a function of galactic coordinates of which we use the inverse of to weight our subsample stars.}
    \label{fig:totalselectionfunctionmean}
\end{figure}

We calculate $S_\mathrm{sub}(\alpha, \delta, G)$ similarly to $S_\mathrm{Gaia}(\alpha, \delta, G)$ by using a $k$-nearest-neighbour scheme. We take the $128$-nearest-neighbours from a composite space of positions-on-the-sky and $G$-band magnitude, i.e. $N_{128}(\alpha, \delta, G)$. However in defining this composite space, we weight the features $(\alpha, \delta, G)$ so that a unit ball now encompasses the area of a HEALPix level 5 pixel while simultaneously spanning $0.2$ mag. This is the same feature weighting as with the binning scheme done by \citet{CastroGinard2023} and in \texttt{GaiaUnlimited} (by default) -- the difference being that in our $k$-nearest-neighbour construction of $S_\mathrm{sub}(\alpha, \delta, G)$, we effectively have smooth and locally adaptive bins that are always centred on the query star and always retain the same relative feature scaling. With this, the subsample selection function is then calculated according to
\begin{equation}
    S_\mathrm{sub}(\alpha, \delta, G) = \frac{|N_{128}(\alpha, \delta, G) \cap \mathrm{sub}|}{128}\, .
\end{equation}
This is simply the fraction of GDR3 stars (with valid positions and $G$-band magnitudes) remaining in a local neighbourhood surrounding the query point, i.e. $(\alpha, \delta, G)$. Since we do not provide the $G$-band magnitude to \astrolink, we also need to marginalise over this feature in Eq.~\ref{eq:totalsf_withG} to get a position-on-the-sky-only dependent total selection function. As such,
\begin{equation} \label{eq:totalsf_marginalisedG}
    S_\mathrm{total}(\alpha, \delta) = \int_\mathrm{Gaia} S_\mathrm{sub}(\alpha, \delta, G)\, S_\mathrm{Gaia}(\alpha, \delta, G)\, p_\mathrm{MW}(G | \alpha, \delta)\, dG
\end{equation}
where $p_\mathrm{MW}(G | \alpha, \delta)$ is the true probability distribution of would-be (were there no selection biases) GDR3 $G$-band magnitudes of stars in the MW at a position-on-the-sky of $(\alpha, \delta)$. This can be estimated from a model of the MW (which we aim to avoid in this analysis) or a deeper catalogue, like DECaPS -- to which we already have a connection via $S_\mathrm{Gaia}(\alpha, \delta, G)$. To be explicit, we have that 
\begin{equation} \label{eq:intrinsicGdistributionproportionality}
    p_\mathrm{MW}(G | \alpha, \delta) \propto \frac{p_\mathrm{Gaia}(G | \alpha, \delta)}{S_\mathrm{Gaia}(\alpha, \delta, G)}\ ,
\end{equation}
which can be normalised to give
\begin{equation} \label{eq:intrinsicGdistribution}
    p_\mathrm{MW}(G | \alpha, \delta) = \frac{\frac{p_\mathrm{Gaia}(G | \alpha, \delta)}{S_\mathrm{Gaia}(\alpha, \delta, G)}}{\int_\mathrm{Gaia} \frac{p_\mathrm{Gaia}(G | \alpha, \delta)}{S_\mathrm{Gaia}(\alpha, \delta, G)}\, dG}\ .
\end{equation}
Combining Eqs.~\ref{eq:totalsf_marginalisedG}~\&~\ref{eq:intrinsicGdistribution} and simplifying we get that
\begin{align}
    S_\mathrm{total}(\alpha, \delta) &= \frac{\int_\mathrm{Gaia} S_\mathrm{sub}(\alpha, \delta, G)\, p_\mathrm{Gaia}(G | \alpha, \delta)\, dG}{\int_\mathrm{Gaia} \frac{p_\mathrm{Gaia}(G | \alpha, \delta)}{S_\mathrm{Gaia}(\alpha, \delta, G)}\, dG} \notag \\
    &= \frac{\mathbb{E}_{G \sim \mathrm{Gaia}}[S_\mathrm{sub}(\alpha, \delta, G)]}{\mathbb{E}_{G \sim \mathrm{Gaia}}[\frac{1}{S_\mathrm{Gaia}(\alpha, \delta, G)}]}\, .
\end{align}
In practice we evaluate these integrals using the $128$-nearest-neighbours on the sky, $N_{128}(\alpha, \delta)$. As this is a local sample of Gaia stars, the expectation values can be estimated as simple means over this set, i.e. 
\begin{align}
    S_\mathrm{total}(\alpha, \delta) &= \frac{\frac{1}{128}\sum_{j \in N_{128}(\alpha, \delta)} S_\mathrm{sub}(\alpha_j, \delta_j, G_j)}{\frac{1}{128}\sum_{j \in N_{128}(\alpha, \delta)} \frac{1}{S_\mathrm{Gaia}(\alpha_j, \delta_j, G_j)}} \notag \\
    &= \frac{\sum_{j \in N_{128}(\alpha, \delta)}\frac{|N_{128}(\alpha_j, \delta_j, G_j) \cap \mathrm{sub}|}{128}}{\sum_{j \in N_{128}(\alpha, \delta)} \frac{1}{S_\mathrm{Gaia}(\alpha_j, \delta_j, G_j)}}\, .
\end{align}
Notice that this is of the form $N_\mathrm{sub}/N_\mathrm{MW}$ with these values being taken from a small region on the sky around the sky-position of $(\alpha, \delta)$ and over the Gaia $G$-band magnitude range. As such, we can extend this to a Bayesian estimate by recognising that it represents the maximum likelihood estimator (MLE) of a binomial probability -- the chance that a star at a given sky location is included in our subsample, given an effective count of $N_\mathrm{sub}$ selected stars out of $N_\mathrm{MW}$ total stars. By modelling this underlying process with a binomial likelihood and placing an uninformative uniform prior -- i.e. Beta(1, 1) -- on the probability, we obtain a posterior distribution of Beta($N_\mathrm{sub} + 1$, $N_\mathrm{MW} - N_\mathrm{sub} + 1$). This posterior represents a distribution over plausible values of $S_\mathrm{total}$, given limited data, and its mean -- $\frac{N_\mathrm{sub} + 1}{N_\mathrm{MW} + 2}$ -- provides a regularised estimate that reduces variance and avoids divergence in low-signal regions\footnote{Instead of using $S_\mathrm{Gaia}(\alpha, \delta, G)$ directly to calculate $N_\mathrm{MW}$, we use $\left(S_\mathrm{Gaia}(\alpha, \delta, G)^2 + \epsilon^2\right)^{1/2}$ so that $1/S_\mathrm{total}(\alpha, \delta)$ does not become extremely large and noisy due to tiny $S_\mathrm{Gaia}(\alpha, \delta, G)$ values. We heuristically set $\epsilon = 1/k_\mathrm{den}$ to reduce noise in the on-sky distribution of $S_\mathrm{total}(\alpha, \delta)$ and temper this effect on the scale of the neighbourhood used by \astrolink\ to calculate local densities.}. This Bayesian formalism is the same as is used by \citet{CastroGinard2023} and is explained in more detail therein. Fig.~\ref{fig:totalselectionfunctionmean} shows how the total selection function mean varies on the sky.


\subsection{Final data space} \label{sec:datatransformation}
We now look to transform the subsample 5d astrometry to a Cartesian coordinate system, as \astrolink\ treats the data as being endowed with at least a locally Euclidean metric. In doing this we assume a zero radial velocity for all stars, which keeps the intrinsic dimensionality of the subsample data set at five but allows for the use of a Euclidean distance on the data. However, even though we now expect our subsample to only contain 5d astrometric solutions with reliable errors, these error values still vary by many orders of magnitude with some solutions being highly inaccurate and even having negative parallaxes.

The largest sources of error produced by the map $(\alpha, \delta, \mu_\alpha, \mu_\delta, r) \mapsto ({\bf x}, {\bf v})$ arise from the photo-geometric distances and proper motion magnitudes. This is not only due to these quantities having larger errors than the positions-on-the-sky, but also since $({\bf x}, {\bf v}) = r\,({\bf\hat x}, \mu{\bf\hat v})$ -- with $\mu = (\mu_\alpha^2 \cos^2\delta + \mu_\delta^2)^{1/2}$ being their proper motion magnitude -- hence they dominate the error propagation. However, while the proper motion error is roughly constant as a function of $\mu$, the distance estimate error scales proportionally to $r^2$. Nevertheless, we can reduce this error and improve the resulting clustering from \astrolink\ -- particularly at large distances where any clustering signal would be otherwise washed out by this uncertainty -- by choosing an appropriate transformation to replace the distance estimate in the conversion to Cartesian coordinates.

\subsubsection{Distance contraction} \label{sec:distancecontraction}
The transformation, $f(r)$, should retain the photo-geometric distance estimate to the degree that it can be trusted, as well as adhere to certain structure preserving conditions. As such, we would like that;
\begin{enumerate}
    \item $f(0) = 0$ and $\frac{df}{dr}(0) = 1$, to preserve continuity and a Euclidean metric closer to the Sun where the distance estimate is more accurate.
    \item $f(r) > 0$ and $\frac{df}{dr} > 0$, to preserve the positivity and order of the estimated distances to stars.
    \item $\Delta f = \frac{df}{dr} \Delta r \propto \frac{df}{dr} r^2$ be constant, to enforce a precision-weighted contracted distance (under first-order error propagation). This is particularly important for large $r$ where the large error would otherwise dominate the error of the Cartesian coordinates.
\end{enumerate}
There are a number of functional forms that satisfy these conditions, but the transformation given by
\begin{equation} \label{eq:distancecontraction}
    f(r) = r_{1/2}\arctan(r/r_{1/2})
\end{equation}
works well in our case. Here $f(r)$ softly approaches a constant value as $r$ becomes large -- effectively painting distant stars onto the sky and reducing the effect of their uncertain distance estimate on the resultant \astrolink\ clustering. Additionally, $\frac{df}{dr}(r_{1/2}) = 1/2$ -- marking the half-way point between a Euclidean- and a zero-dependency of the \astrolink\ clustering on the distance estimate.

We now only need to determine an appropriate value for $r_{1/2}$. If $r_{1/2}$ is too small, then we do not use the distance estimates to the full extent of the information that they provide and the errors from other quantities will dominate despite the reduced error in distance. Conversely, if $r_{1/2}$ is too large then we risk losing clustering signal to large distance estimate errors. To address this issue we choose $r_{1/2}$ to be the median value of $\Delta r / r^2$ that derives from the subsample, which gives $r_{1/2} = \sim139$ pc.

\subsubsection{Relative scaling of positions and velocities}
Since our transformed data is a concatenation of features with differing physical interpretations, we now look to define a meaningful global rescaling that balances the contribution of positions and velocities to the \astrolink\ results. This amounts to normalizing the individual subspaces by some factors so that the Euclidean distance in the rescaled space is equivalent to a feature-weighted Euclidean distance in the unscaled space. We identify three main factors that should contribute to the ideal rescaling:
\begin{enumerate}
    \item \textit{The choice of units.} For example, measuring positions in kpc as opposed to pc (for a given velocity unit) leads to a $1000$-fold difference in the position-dependency of clustering results relative to the velocity-dependency. 
    \item \textit{The size of uncertainties.} If, for example, the uncertainties of the velocities are much larger than those of positions, then the contribution of the velocity estimates of the subsample should be much less than that of positions. 
    \item \textit{The amount of clustering information.} If the positions of stars follow a very multi-modal distribution while their velocities follow a uniform distribution, then the positions subspace holds more clustering information than the velocities subspace. 
\end{enumerate}
Each of these factors is scale-dependent and so a normalization by all three is not possible. We find that normalizing by the median uncertainty size in both subspaces is most effective. This implicitly removes the choice-of-unit dependence while also being the optimal rescaling defined by the Fisher information metric for a data set with homoskedastic uncertainties -- which we have effectively generated by contracting the distance as in Sec.~\ref{sec:distancecontraction}. With this scaling scheme we do not account for the amount of clustering information, however we find that the resultant catalogue cross-match (see Sec.~\ref{sec:catalogue_crossmatching}) is improved if we rescale the velocity values by a factor of two -- suggesting that some additional clustering information lies in the velocity domain relative to the position domain after rescaling by their uncertainties.

\section{Results: The first look} \label{sec:results_first_look}

By applying \astrolink\ to the contracted data space (Sec.~\ref{sec:datatransformation}) of our augmented GDR3 subsample (Sec.~\ref{subsec:cataloguecleaning}) that is weighted by the inverse of the total selection function (Sec.~\ref{subsec:selectionfunction}) and has intrinsic dimensionality of $5$ (Sec.~\ref{subsec:intrinsic_dimensionality}), we obtain a measure of the density-based clustering structure of our galaxy and its surrounds in the form of a binary-tree of overdensities. Each of these groups are potentially astrophysical clusters, although undoubtedly many are simply noisy density fluctuations since together they represent all detectable overdensities within the data space provided. Since \astrolink\ automatically fits a model to the distribution of group prominences (a quantity proportional to the logarithm of the overdensity factor), we are also provided with a statistical significance value for each overdensity.

\begin{figure}
    \centering
    \includegraphics[trim={4mm 5mm 5mm 4mm}, clip, width=\columnwidth]{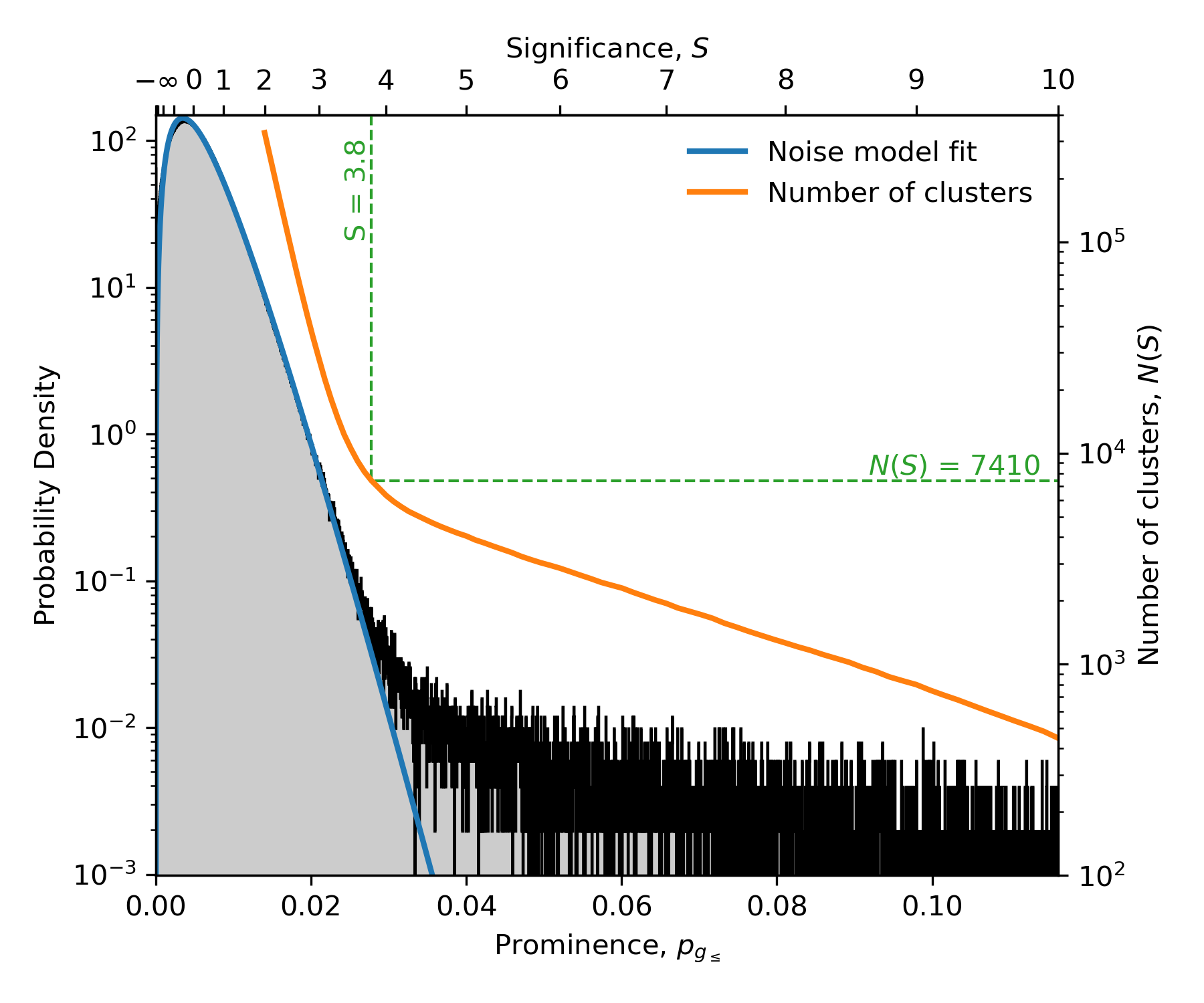}
    \caption{The distribution of subgroup prominences (grey histogram), the fitted noise model (solid blue curve), the relationship between the number of clusters and the significance threshold (solid orange line), and the chosen significance threshold value and the corresponding number of clusters giving $N(3.8)=7410$ (dashed green lines).}
    \label{fig:prominence_model}
\end{figure}

\astrolink\ extracts clusters from this binary-tree by imposing a minimum significance threshold ($S$); above which the overdensities are labelled as \textit{clusters} and below which they can be considered noise. \astrolink\ chooses the value of $S$ automatically by default, however Fig.~\ref{fig:prominence_model} shows that there are two regimes in the relationship between the number of clusters ($N$) and $S$ -- one dominated by noisy density fluctuations and the other by \textit{probably} real astrophysical structures. Each of these regimes represent a different power law relationship between $N$ and the minimum overdensity factor extracted at a given $S$. We choose to extract clusters at a significance threshold of $S=3.8$, as it approximately marks the transition point between these two regimes and lowering $S$ further is expected to result in the incorporation of more noisy groups than astrophysical ones into the cluster hierarchy. Later (Sec.~\ref{sec:catalogue_crossmatching}), we show that this choice is robust as it results in a hierarchy of clusters that compares well to existing catalogues of known structures.

\astrolink\ finds $7410$ clusters under the choice of $S=3.8$. The cluster size distribution is highly skewed, spanning from $3$ to $11,520,440$ members with a median of $62$. In total, there are $42,274,805$ stars found in clusters representing $4.1\%$ of the subsample. The normalised Shannon entropy ($\in [0, 1]$, where the latter would correspond to a uniform assignment across clusters) of the flattened cluster labels provided to each star (they are assigned the label of the smallest cluster they belong to) is $\sim 0.033$ -- representing a highly structured distribution. While about $77\%$, $16\%$, and $5\%$ of clusters existing at the first three levels of the hierarchy respectively, others sit deeper, with the maximum tree depth of the cluster hierarchy reaching level $6$ -- reflecting the intrinsically multi-scale structure of the Milky Way. Even more impressively, \astrolink\ is able to provide this clustering of our GDR3 subsample in $\sim80$ minutes running on 48 CPU cores. This is in stark contrast to the reported runtime of $\sim8$ days taken by HDBSCAN \citep{Campello2015, McInnes2017} when applied to the same feature space of $\sim729.7$ million GDR3 stars using the same number of CPU cores \citep{Hunt2021}.

\begin{figure*}
    \centering
    \includegraphics[trim={7mm 3mm 7mm 3mm}, clip, width=\linewidth]{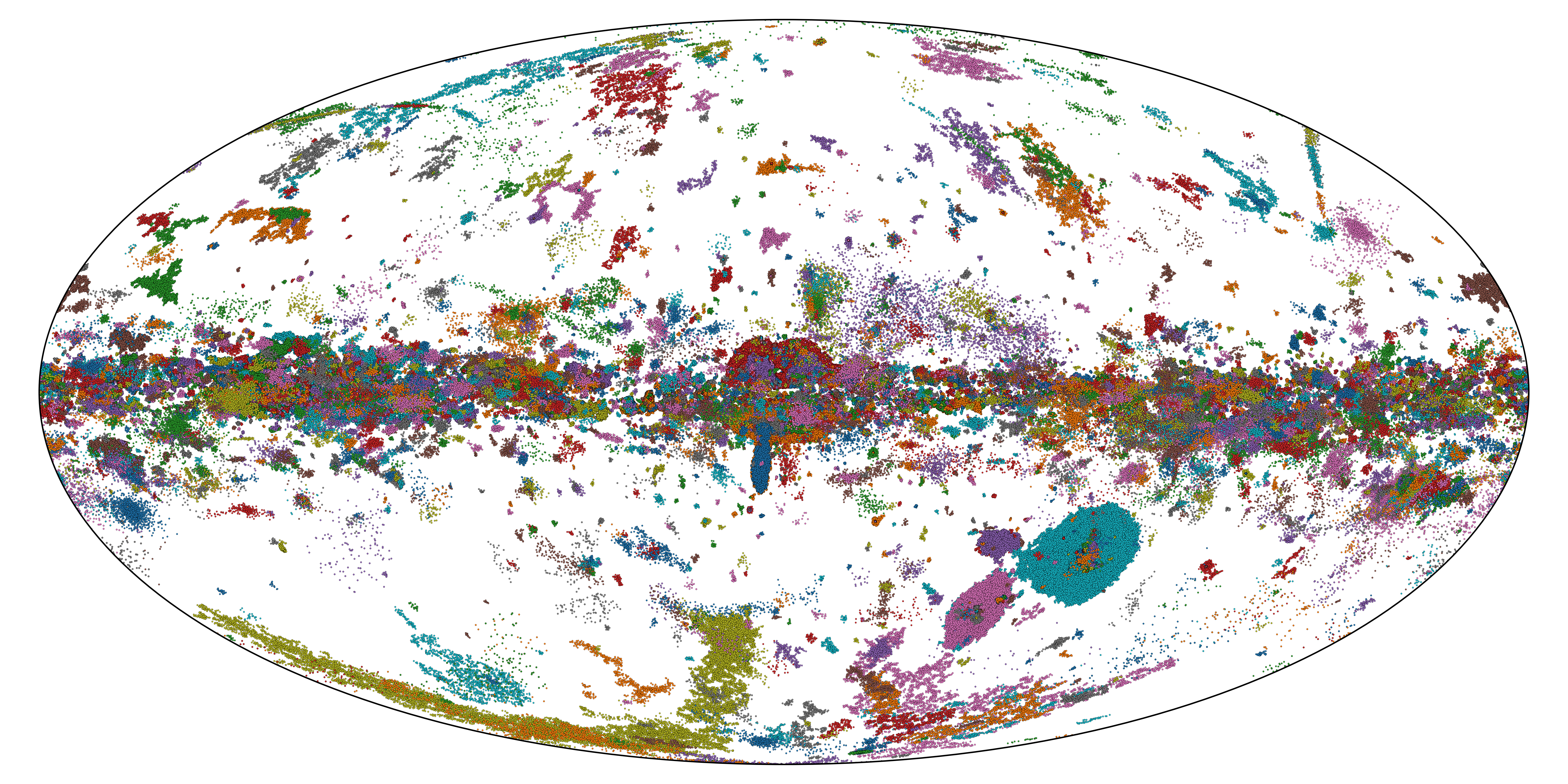}
    \caption{Positions-on-the-sky of stellar members of clusters found by \astrolink\ under the choice of $S=3.8$. Here, the colour of the marker is chosen from a cycle using the cluster label of the smallest cluster in the hierarchy that the star belongs to.}
    \label{fig:all_clusters_on_sky}
\end{figure*}

\begin{figure*}
    \centering
    \includegraphics[trim={7mm 3mm 7mm 3mm}, clip, width=\linewidth]{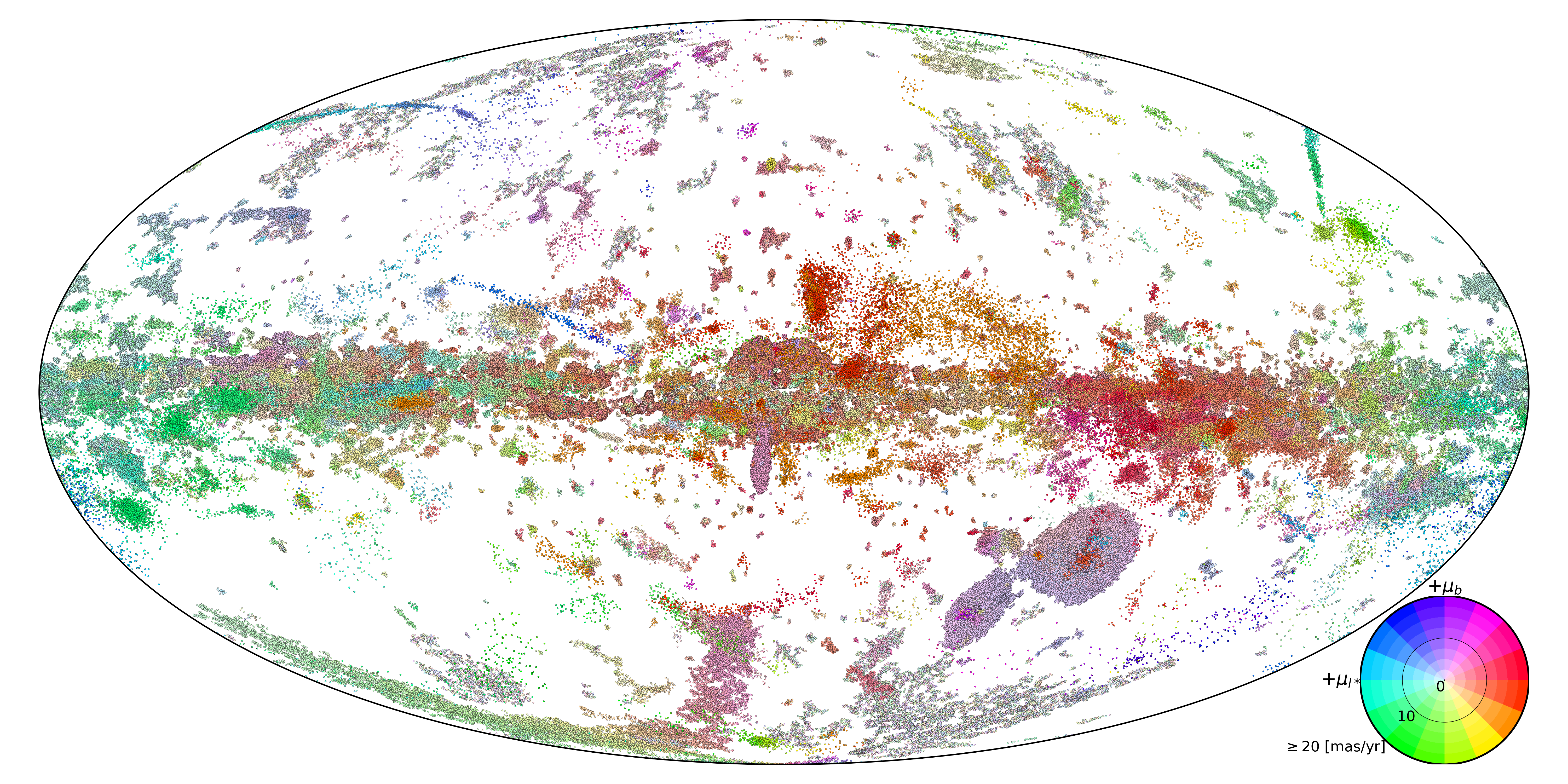}
    \caption{The same as for Fig.~\ref{fig:all_clusters_on_sky}, except here the marker colour is chosen according to the proper motion direction and magnitude of that star. The colour wheel is the lower right panel shows this correspondence.}
    \label{fig:all_clusters_on_sky_proper_motions}
\end{figure*}

Figs.~\ref{fig:all_clusters_on_sky}~\&~\ref{fig:all_clusters_on_sky_proper_motions} depict the positions-on-the-sky in galactic coordinates of all stellar members of all $7410$ clusters. The two figures colour the markers by cluster label and proper motions respectively in order to highlight the structures found. From them, it is easy to see that the clusters found represent a wide range of structure types; from compact groups like open and globular clusters, to diffuse streams and composite/morphological features including parts of the galactic disc and bulge. The spatial distribution reflects both intrinsic Galactic structure and residual selection/transformation effects, e.g. at low Galactic latitudes where extinction effects can be observed or at high Galactic latitudes where the effect of contracting large distances has merged many stars in spatially noisy groups (likely composite structures containing multiple real astrophysical objects, see Sec.~\ref{subsec:cluster_taxonomy}). Many of the visually compact clusters exhibit tight proper motion coherence, while diffuse structures tend to display systematic kinematic gradients consistent with tidally disrupted or large-scale Galactic components.

\section{Analysis: Cross-matching and discovery}  \label{sec:catalogue_crossmatching}

\subsection{Availability-aware comparison metrics} \label{subsec:availability_comparison_metrics}
We now set out to compare the clusters found by \astrolink\ to those that appear in existing catalogues of known structures. Many catalogues report probabilistic memberships, for the purpose of comparison we convert these to binary labels by assigning stars to a cluster if their reported membership probability exceeds $0.5$, thereby harmonising the representation with the binary memberships returned by \astrolink. As such, we opt to use the standard set-overlap metrics of recovery, purity, and Jaccard index -- sometimes called recall/completeness, precision, and Jaccard similarity coefficient respectively. Specifically for two clusters $A$ and $B$, the recovery of $B$ w.r.t. $A$, the purity of $A$ w.r.t. $B$, and the Jaccard index of $A$ and $B$, are defined as
\begin{align} \label{eq:comparison_metrics}
    \mathrm{R}(A \to B) &= \frac{|A \cap B|}{|B|}\, , \notag \\
    \mathrm{P}(B \to A) &= \frac{|A \cap B|}{|A|}\, ,\ \mathrm{and} \notag \\
    \mathrm{J}(A, B) &= \frac{|A \cap B|}{|A \cup B|}\, .
\end{align}
However, in the case that the two clusters have been drawn (via some clustering algorithm) from two different parent sets of data points, these metrics need to be conditioned on that knowledge. So we develop an availability-aware generalisation of the recovery, purity, and Jaccard index. This adjustment reduces the comparison domain to the shared space between the parent sets ($U$). For a given cluster pair, consisting of \astrolink\ cluster $A_i$ and catalogue cluster $C_j$ having been drawn from corresponding parent sets $S_A$ and $S_C$, we derive the availability-aware analogues of Eq.~\ref{eq:comparison_metrics} by first intersecting the input clusters with $U$ and simplifying to get
\begin{align} \label{eq:availability_comparison_metrics}
    \mathrm{R}(A_i \to C_j \mid U) &= \frac{\left| A_i \cap C_j \right|}{\left| C_j \cap U \right|}\, , \notag \\
    \mathrm{P}(C_j \to A_i \mid U) &= \frac{\left| A_i \cap C_j \right|}{\left| A_i \cap U \right|}\, ,\ \mathrm{and} \notag \\
    \mathrm{J}(A_i, C_j \mid U) &= \frac{\left| A_i \cap C_j \right|}{\left| (A_i \cup C_j) \cap U \right|},\ \mathrm{where}\ U=S_A \cap S_C\, .
\end{align}
This formulation is exact provided the parent sets used to find each catalogue of clusters are known. In practice however, the effective parent set of stars ($S_C$) from which a catalogue of clusters is found, is rarely provided alongside the cluster catalogue itself (as is the case with the catalogues we compare to, Secs.~\ref{subsubsec:hunt2024}--\ref{subsubsec:galstreams}), which introduces uncertainty in the comparison domain.

So instead, we compute the limiting bounds on the effective universe $U$ such that; (1) $S_C$ is any superset of $S_A$ giving $U=S_A$, and (2) $S_C$ is the union of all stars appearing in the catalogue giving $U = S_A \cap (\bigcup_j C_j)$. These bounds yield minimum and maximum values respectively, for both the purity of $A_i$ w.r.t. $C_j$ and for the Jaccard index of $A_i$ and $C_j$. Intuitively, the lower bound assumes the comparison catalogue has searched all stars in $S_A$, thereby treating every unmatched star in $A_i$ as a potential false positive, whereas the upper bound assumes the catalogue only searched stars that appear in one of its reported clusters, excluding all other unmatched stars from the comparison. The true values of these quantities are guaranteed to be in this range and the bounds simply encapsulate our lack of knowledge about the \textit{true negatives} w.r.t. each catalogue. The recovery of $C_j$ w.r.t. $A_i$ is unaffected by the choice of $S_C$ since $C_j$ is always its subset under these two scenarios.

We use the availability-aware comparison metrics in two complementary ways. First, we evaluate how well each catalogue is matched as a function of the \astrolink\ significance threshold $S$, by tracking for each catalogue cluster its best-matching \astrolink\ leaf cluster at each $S$ as detailed in Sec.~\ref{subsec:catalogues} -- enabling a performance assessment that is independent of our choice of $S=3.8$ made in Sec.~\ref{sec:results_first_look}. Second, at the adopted optimal threshold, we construct a bi-directional best-match relation between \astrolink\ and catalogue clusters, which forms the basis of the structural classifications presented in Sec.~\ref{subsec:cluster_taxonomy}.

\subsection{Catalogues crossmatching} \label{subsec:catalogues}

In this section we analyse the degree to which different catalogues are matched by the \astrolink\ clusters (discussed in Sec.~\ref{sec:results_first_look}) via the pairwise availability-aware comparison metrics (discussed in Sec.~\ref{subsec:availability_comparison_metrics}). The full set of pairwise comparisons defines a multi-weighted bipartite graph between the \astrolink\ and catalogue clusters. While this graph encodes rich structural information, it is difficult to summarise in a compact and interpretable way. For the purposes of global performance assessment, we therefore reduce this graph to a single \textit{best-match} relation for each catalogue cluster. In general, the best-match \astrolink\ cluster ($A_{i^\star}$) is defined such that
\begin{align}
    i^\star(j; S) &= \arg\max_i \, M(A_i, C_j \mid U)
\end{align}
where $M$ is a metric chosen from among the availability-aware comparison metrics. This reduction does not enforce one-to-one correspondences, and as such preserves the possibility of fragmentation and merging -- which we analyse separately in Sec.~\ref{subsec:cluster_taxonomy}.

The choice of $M$ is made on a per-catalogue basis -- see the dedicated sub-subsection for each catalogue (Secs.~\ref{subsubsec:hunt2024}--\ref{subsubsec:galstreams}) for more details. In general, the Jaccard index is the correct choice if the catalogue is considered complete, but since this is symmetric, it becomes the wrong choice when the catalogue provides under- or over-confident membership lists. Additionally, the lower or upper bound may be preferred if the catalogue is expected to have been drawn from an effective GDR3 stellar sample that is comparable to the one we use in this work or that is much smaller than that, respectively. Finally, since many catalogue clusters are only partially represented in our subsample, raw comparison metric values may be noisy. We therefore compute the coverage fraction of each catalogue cluster by our subsample, $F_j$, defined as the fraction of its stars that appear in $S_A$ -- and weight all summary statistics accordingly. This prevents weakly-covered clusters from dominating averages. To be explicit and for each catalogue, we compute summary statistics
\begin{align}
    \tilde{M} &= \frac{\sum_i F_{j^\star} \cdot M(A_i, C_{j^\star} \mid U)}{\sum_i F_{j^\star}}\quad \mathrm{and} \notag \\
    \mathrm{Matched\ Fraction} &= \frac{\sum_i F_{j^\star} \cdot {\bf 1}\left(M(A_i, C_{j^\star} \mid U) > 0.5\right)}{\sum_i F_{j^\star}}.
\end{align}
Here ${j^\star}$ denotes the best-matching catalogue cluster under the adopted metric, see Eq.~\ref{eq:bestmatchdefinition} and surrounding text for a more thorough definition. For each catalogue (and choice of $M$) we show these summary statistics as a function of the \astrolink\ significance threshold ($S \in [2, 10]$) in Figs.~\ref{fig:Hunt2024_crossmatch_per_significance}--\ref{fig:galstreams_crossmatch_per_significance}, which independently reveals the quality of our choice of $S=3.8$ but also indicates whether \astrolink\ typically fragments or merges structures of given type.

\subsubsection{Hunt \& Reffert (2024)} \label{subsubsec:hunt2024}
The catalogue of \citet{Hunt2024} presents a homogeneous all-sky search for stellar clusters in GDR3 using the HDBSCAN density-based clustering algorithm \citep{Campello2015, McInnes2017}. The parent sample comprises approximately 
$729.7$ million stars selected from GDR3 after applying a neural-network-based quality filter to remove sources with unreliable astrometry. The resulting catalogue contains 7167 clusters with full membership lists and derived parameters, including 4782 previously known systems, 2387 new candidates, and 134 known globular clusters. For each cluster, the authors provide astrometric and astrophysical parameters, membership probabilities, signal-to-noise ratio estimates, and structure-type classifications (open clusters, moving groups, globular clusters, “too distant”, and “rejected”). As the largest HDBSCAN-based cluster search on GDR3 with probabilistic memberships, this catalogue provides a natural primary benchmark for assessing the performance of \astrolink.

\begin{figure}
    \centering
    \includegraphics[trim={3mm 4mm 3mm 3mm}, clip, width=\columnwidth]{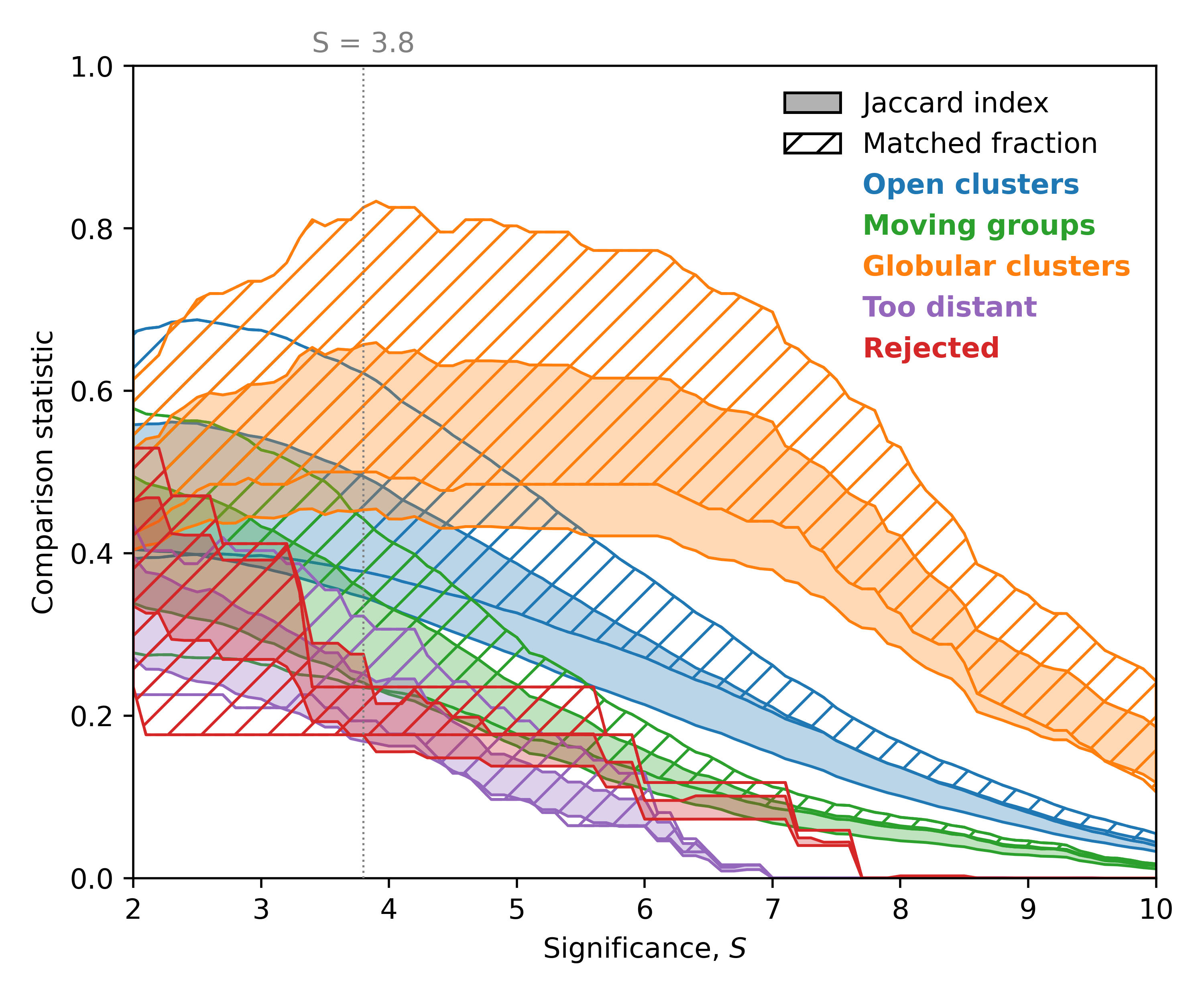}
    \caption{Crossmatch performance between \astrolink\ clusters and the \citet{Hunt2024} catalogue as a function of the \astrolink\ significance threshold, $S$. Solid curves show the coverage-weighted mean availability-aware Jaccard index of the best-matching \astrolink\ cluster for each catalogue cluster, with shaded regions indicating the bounds arising from uncertainty in the effective comparison domain. Hatched regions show the fraction of catalogue clusters with Jaccard index $>0.5$ between the same comparison domain bounds. Colours denote cluster type (open clusters, moving groups, globular clusters, too distant, and rejected). The vertical dotted line marks the adopted optimal threshold of $S=3.8$.}
    \label{fig:Hunt2024_crossmatch_per_significance}
\end{figure}

Given the very large parent sample used by \citet{Hunt2024}, we adopt the lower bound of the availability-aware Jaccard index when defining the best-matching \astrolink\ cluster, corresponding to the assumption that their effective parent set is broadly comparable to our own, which contains $\sim41\%$ more stars. We examine the coverage-weighted mean Jaccard index ($\tilde{J}$) and the fraction of clusters with $J>0.5$ as functions of the \astrolink\ significance threshold $S$, separately for each structure type in the catalogue as depicted in Fig.~\ref{fig:Hunt2024_crossmatch_per_significance}. Open clusters and globular clusters exhibit clear maxima in Jaccard index as a function of $S$, peaking near $S\approx3$ and $S\approx4$, respectively, while moving groups and lower-confidence classes show a monotonic increase towards lower $S$. At the adopted threshold $S=3.8$, globular clusters retain high agreement ($\tilde{J} \in [0.5, 0.65]$), whereas open clusters show slightly reduced overlap relative to their peak and moving groups remain comparatively low. This behaviour suggests that, relative to the Hunt \& Reffert classifications, \astrolink\ leaf clusters at $S=3.8$ may only mildly fragment compact globular clusters (if at all), while tending to merge or recover with lower purity (likely most often the latter, see that Tab.~\ref{tab:hierarchy_relationship_classification} still shows many true one-to-one matches) more diffuse systems such as open clusters and moving groups.

\subsubsection{The Unified Cluster Catalogue} \label{subsubsec:ucc}
The Unified Cluster Catalogue \citep{Perren2023} (UCC) compiles essentially all published Milky Way open clusters into a single homogenized database. Its initial release contained nearly 14\,000 unique open clusters, increasing to 18\,299 in the version used here (version 260615), far exceeding the size of traditional catalogues such as WEBDA \citep{Mermilliod2003} or SIMBAD \citep{Wenger2000}. Each cluster is re-analysed using \textit{Gaia} DR3 photometry and astrometry down to $G \approx 20$ with the probabilistic membership algorithm \texttt{fastMP} \citep{Perren2023}, resulting in more than one million stars assigned cluster membership probabilities. The UCC further introduces a two-letter quality classification, where the first letter quantifies astrometric coherence and the second photometric consistency. Each letter ranges from A (highest quality) to D (lowest), producing combined classes such as AA or DC. Approximately 40\% of clusters are assigned high-confidence classifications (AA, AB, or BA), while roughly 15\% fall into the lowest-quality categories. The catalogue is publicly accessible and provides a unified, quality-assessed reference for the open cluster population of the Milky Way.

\begin{figure}
    \centering
    \includegraphics[trim={3mm 4mm 3mm 3mm}, clip, width=\columnwidth]{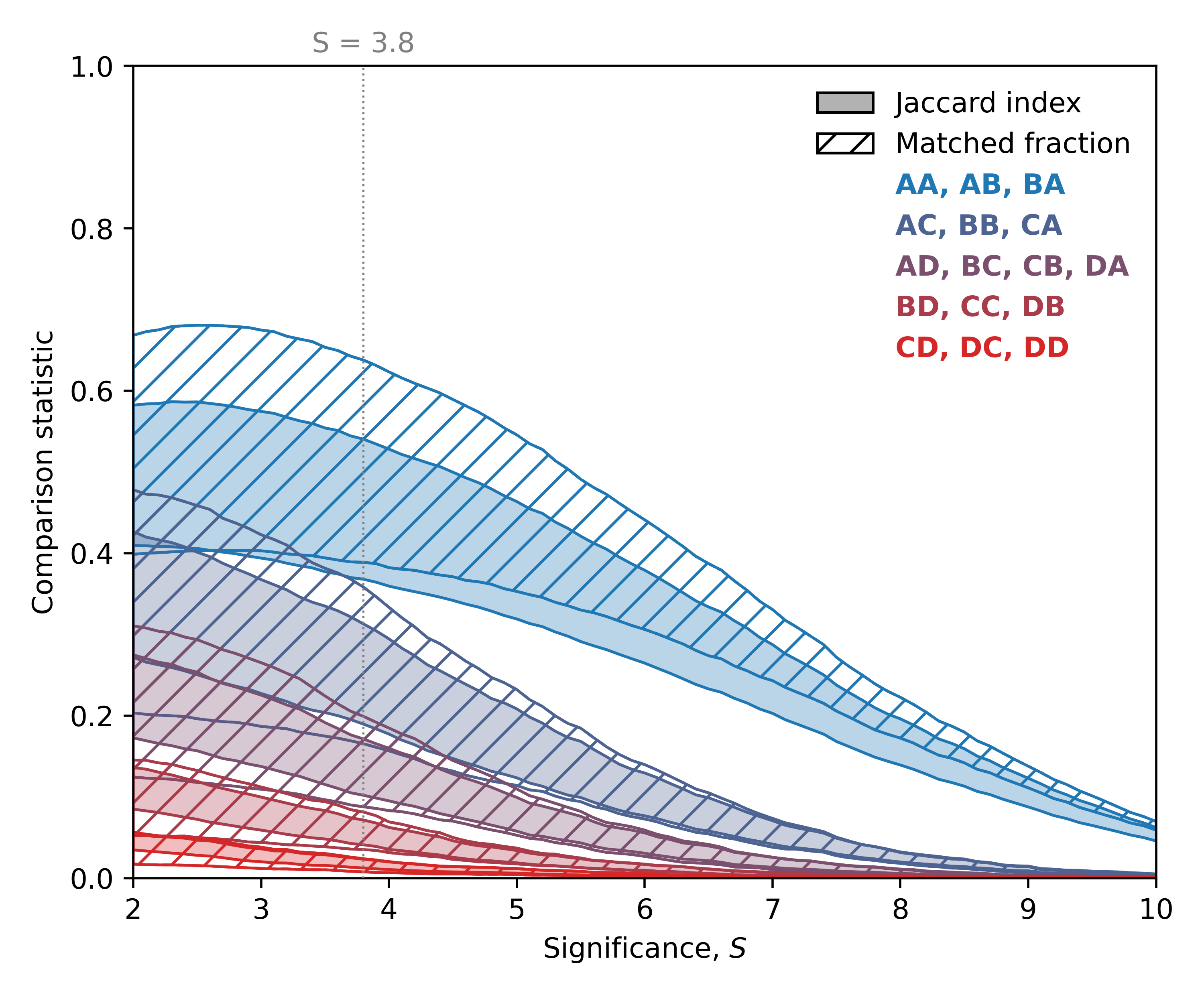}
    \caption{Crossmatch performance between \astrolink\ clusters and the UCC \citep{Perren2023} as a function of the \astrolink\ significance threshold, $S$. Solid curves show the coverage-weighted mean availability-aware Jaccard index of the best-matching \astrolink\ cluster for each UCC cluster, with shaded regions indicating bounds arising from uncertainty in the effective comparison domain. Hatched regions show the fraction of catalogue clusters with Jaccard index $>0.5$ between the same comparison domain bounds. Colours denote UCC quality class groups. The vertical dotted line marks the adopted optimal threshold.}
    \label{fig:UCC_crossmatch_per_significance}
\end{figure}

Given that the UCC aggregates clusters identified across the literature, many of which were derived from large \textit{Gaia} DR3 samples comparable to our own, we adopt the lower bound of the availability-aware Jaccard index to define the best-matching \astrolink\ cluster. We examine both the coverage-weighted Jaccard index ($\tilde{J}$) and the matched fraction ($N_{\tilde{J} > 0.5} / N_\mathrm{total}$) as functions of $S$, subdividing the sample into combinations of the UCC cluster quality classes (\{AA, AB, BA\}, \{AC, BB, CA\}, \{AD, BC, CB, DA\}, \{BD, CC, DB\}, \{CD, DC, DD\}) as shown in Fig.~\ref{fig:UCC_crossmatch_per_significance}. The quality classes are a two-character indication of spatial and photometric consistency with being an open cluster. A clear trend is observed in which the peak of the $\tilde{J}$ distribution shifts toward lower $S$ values and lower peak amplitudes as the UCC quality class decreases. For the highest-quality classes, the peak occurs at $S \approx 2.5$, where $\tilde{J}$ lies in the range $[0.4,\,0.6]$ and the matched fraction in $[0.4,\,0.7]$. At our adopted choice of $S=3.8$, $\tilde{J}$ for these high-quality clusters remains in the range $[0.4,\,0.55]$, with matched fractions of $[0.4,\,0.65]$. As found for \citet{Hunt2024}, open clusters in the UCC (to which the former contributes a significant portion) are therefore frequently merged or recovered with reduced purity by \astrolink\ leaf clusters at $S=3.8$.

\subsubsection{Vasiliev \& Baumgardt (2021)} \label{subsubsec:vasiliev2021}
The globular cluster catalogue of \citet{Vasiliev2021a, Vasiliev2021b} provides an up-to-date census of all known Milky Way globular clusters based on \textit{Gaia} EDR3 astrometry. The authors determine mean parallaxes and proper motions for approximately 170 clusters, refining their distances, systemic motions, and orbital trajectories. Using these measurements, they derive detailed internal kinematic profiles, including velocity dispersions, rotation signatures, and anisotropy constraints, thereby calibrating the astrometric properties of globular clusters within the \textit{Gaia} reference frame. The catalogue also provides membership classifications for stars in each cluster, with membership lists and analysis tools made publicly available. It therefore constitutes the standard \textit{Gaia}-based reference for the kinematics, orbital properties, and membership structures of Galactic globular clusters.

\begin{figure}
    \centering
    \includegraphics[trim={3mm 4mm 3mm 3mm}, clip, width=\columnwidth]{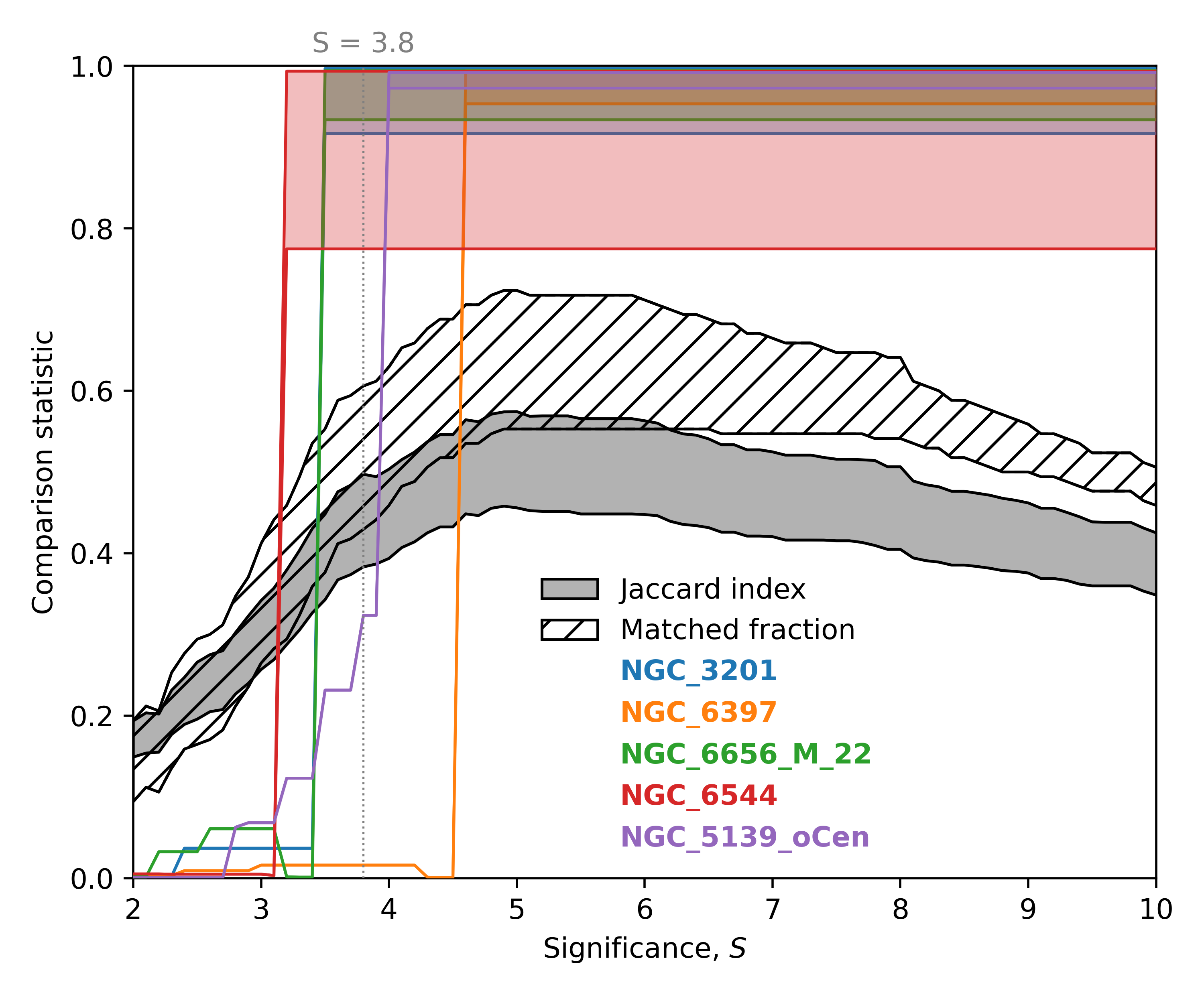}
    \caption{Crossmatch performance between \astrolink\ clusters and the \citet{Vasiliev2021a, Vasiliev2021b} catalogue as a function of the \astrolink\ significance threshold, $S$. Solid curves show the coverage-weighted mean availability-aware Jaccard index of the best-matching \astrolink\ cluster for each globular cluster, with shaded regions indicating bounds arising from uncertainty in the effective comparison domain. Hatched regions show the fraction of catalogue clusters with Jaccard index $>0.5$ between the same comparison domain bounds. Colours denote the Jaccard index of specific globular clusters. The vertical dotted line marks the adopted optimal threshold.}
    \label{fig:Vasiliev2021_crossmatch_per_significance}
\end{figure}

As this catalogue represents a refined compilation of well-established globular clusters -- many identified prior to the \textit{Gaia} era -- we adopt the upper bound of the availability-aware Jaccard index to define the best-matching \astrolink\ cluster. We examine the ranges of the coverage-weighted Jaccard index ($\tilde{J}$) and matched fraction as functions of $S$ for the full sample as shown in Fig.~\ref{fig:Vasiliev2021_crossmatch_per_significance}. The peak correspondence occurs at $S \approx 4.9$, where $\tilde{J} \in [0.45,\,0.575]$ and the matched fraction lies in $[0.55,\,0.75{+}]$. At our adopted choice of $S=3.8$, these ranges decrease to $\tilde{J} \in [\sim0.375,\,0.5]$ and matched fraction in $[0.425,\,0.6]$. For several well-matched examples (NGC\,3201, NGC\,6397, M\,22, NGC\,6544, and Omega Centauri), the Jaccard index range exceeds 0.8 or 0.9 above threshold values of $S$, indicating that \astrolink\ is capable of near one-to-one recovery. However, at $S=3.8$, only NGC\,3201, M\,22, and NGC\,6544 are recovered without clear signs of fragmentation or merging, while NGC\,6397 and Omega Centauri appear fragmented at the \astrolink\ leaf level. This does not preclude a clean match at a higher hierarchical level, but suggests that \astrolink\ may preferentially isolate the cluster core or identify significant internal substructure in these systems. Consistent with the behaviour observed for globular clusters in Hunt \& Reffert (2024), the Vasiliev \& Baumgardt (2021) clusters are therefore sometimes fragmented by \astrolink, a conclusion that is strongly reinforced by Table~\ref{tab:hierarchy_relationship_classification} which shows that many nominal one-to-one matches exhibit additional internal substructure within the \astrolink\ hierarchy.

\subsubsection{Battaglia et al. (2021)} \label{subsubsec:battaglia2021}
The catalogue of Milky Way dwarf galaxies compiled by \citet{Battaglia2021} provides systemic proper motions and membership selections for 74 dwarf galaxies and dwarf galaxy candidates, excluding the Magellanic Clouds. Their analysis combines \textit{Gaia} EDR3 astrometry with a Bayesian membership inference scheme to assign probabilities to stars in each system. The resulting dataset delivers high-quality systemic motions for 66 dwarfs, extending to distances of $\sim1.4\,\mathrm{Mpc}$ and including many recently discovered ultra-faint satellites. The modelling explicitly accounts for the dynamical influence of the Large Magellanic Cloud when deriving satellite orbits. This catalogue therefore constitutes the definitive \textit{Gaia}-based reference for the kinematics and stellar memberships of Milky Way dwarf galaxies.

\begin{figure}
    \centering
    \includegraphics[trim={3mm 4mm 3mm 3mm}, clip, width=\columnwidth]{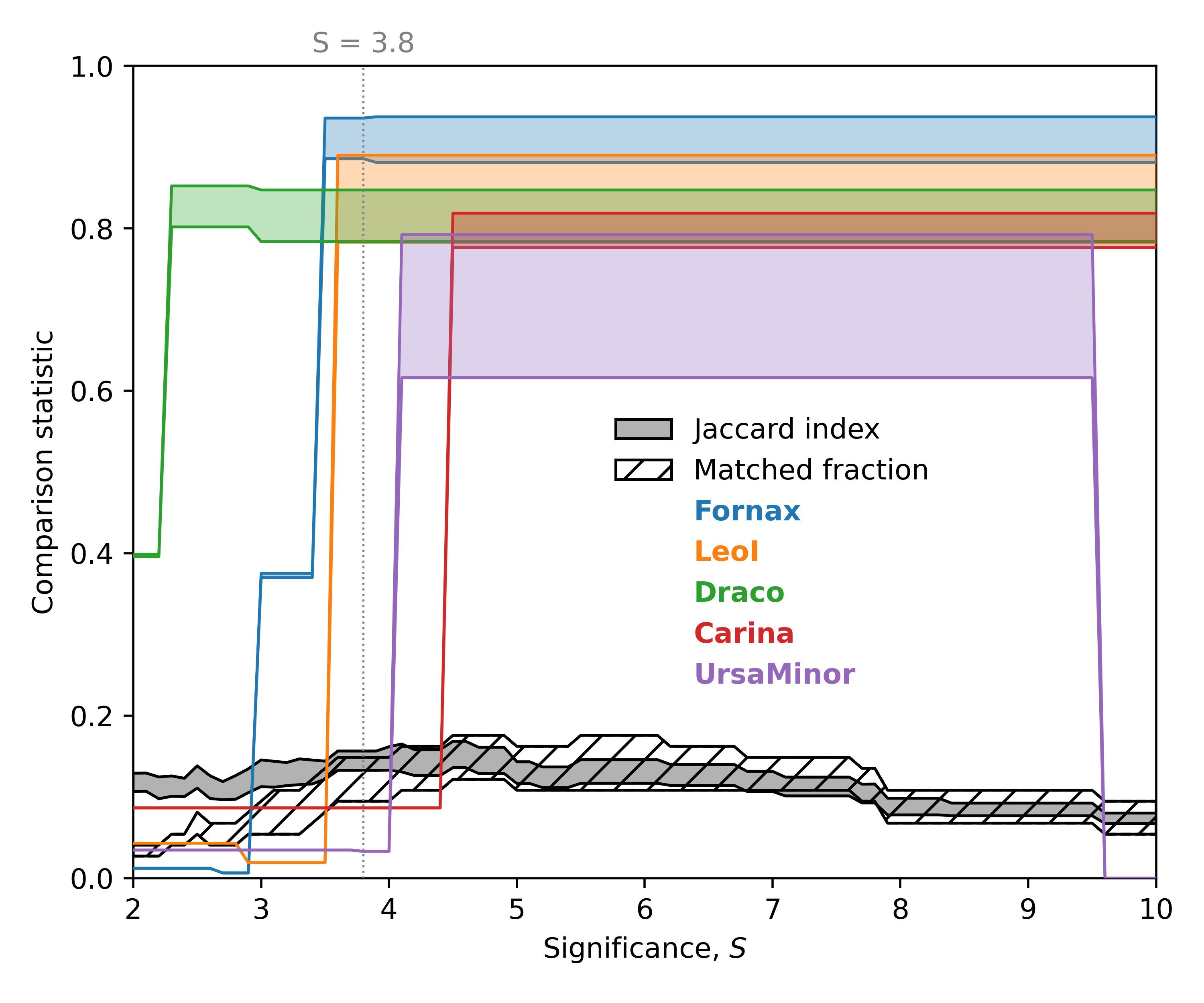}
    \caption{Crossmatch performance between \astrolink\ clusters and the \citet{Battaglia2021} catalogue as a function of the \astrolink\ significance threshold, $S$. Solid curves show the coverage-weighted mean availability-aware Jaccard index of the best-matching \astrolink\ cluster for each dwarf galaxy, with shaded regions indicating bounds arising from uncertainty in the effective comparison domain. Hatched regions show the fraction of catalogue clusters with Jaccard index $>0.5$ between the same comparison domain bounds. Colours denote the Jaccard index of specific dwarf galaxies. The vertical dotted line marks the adopted optimal threshold.}
    \label{fig:Battaglia2021_crossmatch_per_significance}
\end{figure}

As with \citet{Vasiliev2021a, Vasiliev2021b}, this catalogue provides \textit{Gaia}-based membership lists for systems that were largely identified prior to the \textit{Gaia} era or with external surveys. We therefore adopt the upper bound of the availability-aware Jaccard index when defining the best-matching \astrolink\ cluster, corresponding to the assumption that the effective comparison set is best approximated by the full catalogue membership ($S_{\mathrm{C}} = (\bigcup_j C_j)$) rather than $S_{\mathrm{C}} = S_{\mathrm{A}}$. We examine the ranges of the coverage-weighted Jaccard index ($\tilde{J}$) and matched fraction as functions of $S$ for the full dwarf galaxy sample as depicted in Fig.~\ref{fig:Battaglia2021_crossmatch_per_significance}. The peak correspondence occurs at $S \approx 4.5$, where $\tilde{J} \in [0.125,\,0.175]$ and the matched fraction lies in $[0.1,\,0.175{+}]$. At our adopted choice of $S=3.8$, the $\tilde{J}$ range is similar, while the matched fraction decreases slightly. For several well-matched examples (Fornax, Leo~I, Draco, Carina, and Ursa~Minor), the standard Jaccard index exceeds $0.6$ or $0.8$ for most $S \gtrsim 4.5$, demonstrating that \astrolink\ can recover these systems with high fidelity at appropriate scales. At $S=3.8$, however, only Fornax, Leo~I, and Draco are recovered without clear evidence of fragmentation or merging, whereas Carina and Ursa~Minor are likely fragmented or recovered by their core at the leaf level. As for the globular clusters, this does not exclude the possibility of cleaner matches at higher hierarchical levels, but it does suggest that \astrolink\ may preferentially isolate the dense cores of these galaxies or identify internal substructure within them. While this behaviour is consistent with the hierarchy relationships summarised in Table~\ref{tab:hierarchy_relationship_classification}, the typically low coverage of dwarf galaxies within our subsample -- a consequence of our selection function -- means that the evidence for genuine internal substructure is less conclusive than in the case of globular clusters.

\subsubsection{\texttt{galstreams}} \label{subsubsec:galstreams}
For stellar streams, we use the \texttt{galstreams} Python package \citep{Mateu2023}, which compiles a homogenized collection of stream tracks from the literature. Each stream is represented by one or more tracks defined in stream-aligned coordinates, specifying the central locus of the stream on the sky (the $\phi_1$--$\phi_2$ plane) together with any available information on line-of-sight distance, proper motion, and radial velocity as functions of $\phi_1$. As of version~1.0.2, the package contains 217 track segments corresponding to 141 distinct stellar streams, and includes estimates of positional and kinematic widths where available. The \texttt{galstreams} models therefore provide a uniform representation of known tidal streams against which we can compare the \astrolink\ hierarchy. In practice, we use these models to identify all \textit{Gaia} DR3 stars in our subsample that fall within the positional and kinematic footprint defined for each stream.

\begin{figure}
    \centering
    \includegraphics[trim={3mm 4mm 3mm 3mm}, clip, width=\columnwidth]{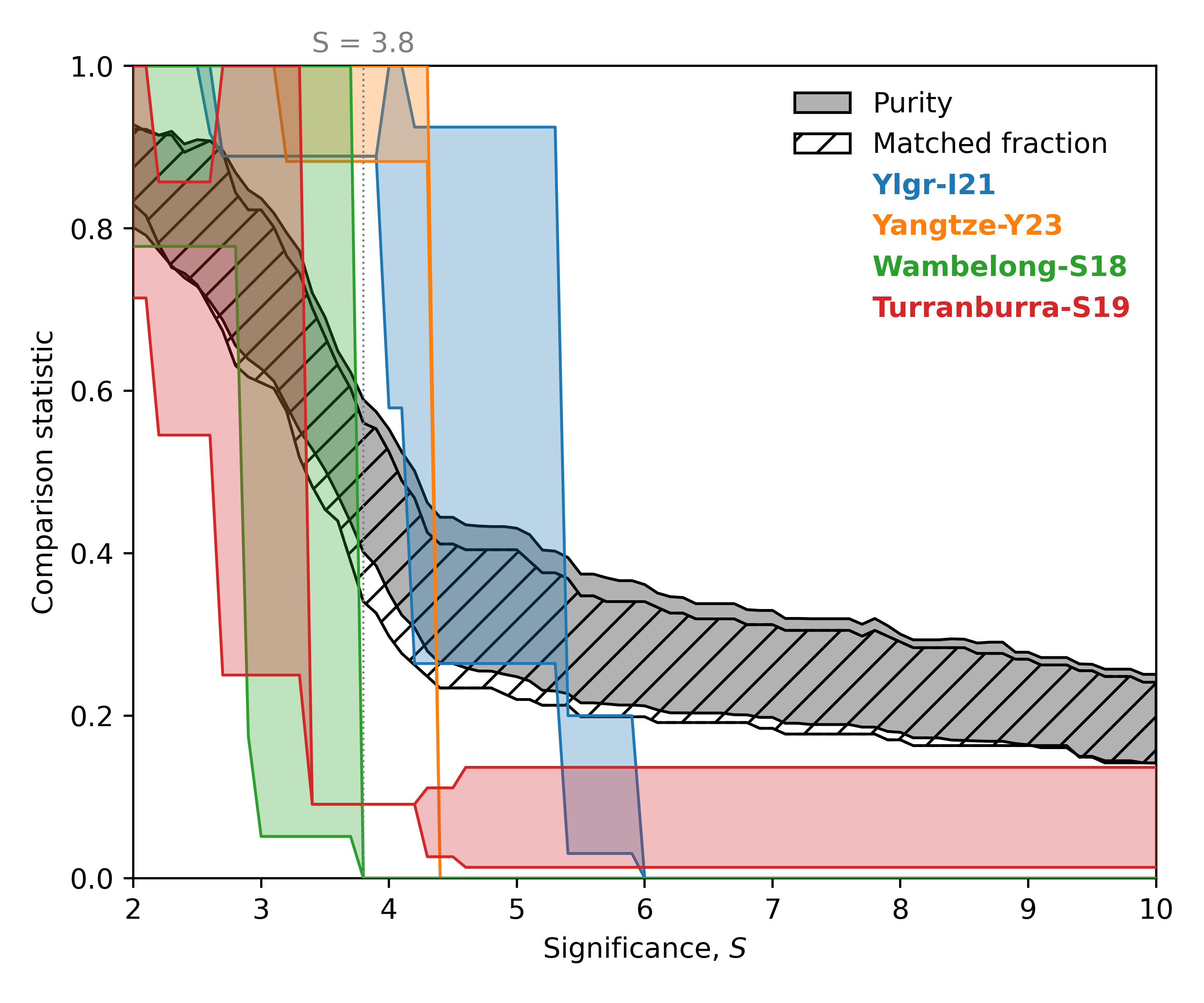}
    \caption{Crossmatch performance between \astrolink\ clusters and the \texttt{galstreams} \citep{Mateu2023} catalogue as a function of the \astrolink\ significance threshold, $S$. Solid curves show the coverage-weighted mean availability-aware purity of the best-matching \astrolink\ cluster for each stream w.r.t the stream itself, with shaded regions indicating bounds arising from uncertainty in the effective comparison domain. Hatched regions show the fraction of catalogue clusters with purity $>0.5$ between the same comparison domain bounds. Colours denote the purity w.r.t specific streams. The vertical dotted line marks the adopted optimal threshold.}
    \label{fig:galstreams_crossmatch_per_significance}
\end{figure}

Unlike the cluster catalogues considered above, the \texttt{galstreams} tracks do not provide statistically validated membership lists, but rather define regions of phase space in which stream members are expected to reside. These regions are necessarily under-specified relative to the true, dynamically cold and coherent tidal structures they represent. While this approach enables the identification of candidate members, it does not provide a reliable characterisation of true non-members, leading to an inherently one-sided estimate of stream membership that is typically contaminated by background stars. Consequently, symmetric set-overlap metrics such as the Jaccard index are not well suited to this comparison. Instead, we adopt the lower bound of the purity as an asymmetric measure to identify \emph{potentially} matching \astrolink\ clusters, recognising that we can confidently exclude certain clusters from the potential match candidates but cannot assert definitive stream classification among the \astrolink\ clusters. This asymmetry has two important implications: some \astrolink\ clusters that overlap stream footprints may in fact correspond more closely to other types of structure, and conversely some genuine stream detections may not be classified as well-matched under this metric. The impact of this ambiguity on the global catalogue classification is discussed further in Sec.~\ref{subsec:cluster_taxonomy}.

We examine the ranges of the coverage-averaged purity ($\tilde{P}$) and the matched fraction -- defined here as the fraction of streams whose highest-purity \astrolink\ leaf cluster has $P > 0.5$ with respect to the stream footprint -- as functions of $S$ for the full stream sample as shown in Fig.~\ref{fig:galstreams_crossmatch_per_significance}. Owing to the asymmetric definition of purity, both $\tilde{P}$ and the matched fraction are expected to decrease monotonically with increasing $S$, as larger leaf clusters naturally dilute purity with respect to the stream footprint; this behaviour is observed in Fig.~\ref{fig:galstreams_crossmatch_per_significance}. Nevertheless, an elbow in $\tilde{P}$ is visible around $S \approx 4.5$, where $\tilde{P} \in [0.25,\,0.45]$, suggestive of a scale at which streams may be more coherently represented, although no firm conclusion can be drawn. Individual examples such as Ylgr, Yangtze, Wambelong, and Turranburra attain the highest purity values of the catalogue, yet visual inspection of Figs.~\ref{fig:all_clusters_on_sky}~\&~\ref{fig:all_clusters_on_sky_proper_motions} indicates that some apparent matches at $S=3.8$ arise from compact clusters lying within stream footprints rather than extended tidal structures. Conversely and from those same figures, well-known streams such as GD-1, Sagittarius, Phlegethon, Hrid, Pal~5, and Gaia-1 appear to be at least partially recovered by \astrolink, even if not classified as high-purity matches under our adopted metric. Taken together, these results suggest that at $S=3.8$ known streams may be modestly fragmented into multiple leaf clusters, while slightly larger $S$ values could enhance their coherence. However, given the asymmetric and footprint-based nature of the comparison, the present analysis cannot conclusively determine whether such fragmentation reflects incomplete recovery or physically meaningful segmentation of progenitors and internal substructure. Although we might reasonably expect the latter as \astrolink\ has been shown to characterise streams in this way in simulated MW-like data \citep{Oliver2024b}.

\subsection{Model-agnostic cluster taxonomy} \label{subsec:cluster_taxonomy}
In Sec.~\ref{subsec:catalogues} we reduced the multi-weighted bipartite graph between the \astrolink\ and catalogue clusters to a catalogue-centric best-match relation $i^*(j; S)$, enabling a performance assessment as a function of the significance threshold $S$. We now fix $S = 3.8$ and use the same availability-aware comparison metrics to construct a \emph{bi-directional} best-match relation. This transforms the performance-oriented analysis into a structural classification of the \astrolink\ hierarchy. Here, we consider all \astrolink\ clusters (excluding the background root), not only leaf clusters. This allows higher-level nodes to match extended or composite systems that are not well represented by individual leaves. For the set of \astrolink\ clusters $\mathcal{A} = \{A_i\}$ and a given catalogue $\mathcal{C} = \{C_j\}$, we define
\begin{align} \label{eq:bestmatchdefinition}
    i^\star(j) &= \arg\max_i \, M(A_i, C_j \mid U), \notag \\
    j^\star(i) &= \arg\max_j \, M(A_i, C_j \mid U),
\end{align}
where $M$ is the catalogue-specific comparison metric defined in Sec.~\ref{subsec:catalogues}. All availability-aware comparison metrics are retained for subsequent interpretation. Because the maximisation is performed independently in each direction, the resulting best-match relation is not, in general, symmetric. This asymmetry encodes structural information about fragmentation and merging.

\subsubsection{Structural relationship classes} \label{subsubsec:structural_relationship_classes}
For each catalogue, we classify relationships using reciprocal best matches and multiplicities of assignment.

\paragraph{One-to-one (reciprocal) match.}
$A_i$ and $C_j$ form a reciprocal pair if
\begin{equation}
    j^\star(i) = j \quad \text{and} \quad i^\star(j) = i.
\end{equation}
These correspond to mutually optimal associations under the adopted metric.

\paragraph{Fragmentation (catalogue perspective).}
A catalogue cluster $C_j$ is fragmented (by \astrolink) if
\begin{equation}
    \left| \{\, i : j^\star(i) = j \,\} \right| > 1,
\end{equation}
i.e. if multiple \astrolink\ clusters identify $C_j$ as their best match. In this case, a single catalogue structure is decomposed within the \astrolink\ hierarchy.

\paragraph{Merging (catalogue perspective).}
An \astrolink\ cluster $A_i$ is merged (by a catalogue) if
\begin{equation}
    \left| \{\, j : i^\star(j) = i \,\} \right| > 1,
\end{equation}
i.e. if multiple catalogue clusters identify $A_i$ as their best match. In this case, a single \astrolink\ cluster combines structures from the catalogue it is being compared to.

\paragraph{Isolated match.}
A cluster is isolated if it has a valid best match, but the match is; not reciprocal, not fragmented; and not merged. Such systems represent hierarchical and structural mismatches.

\paragraph{Unmatched objects.}
Objects for which no valid best match is defined are classified as unmatched. This occurs when there are no members of a cluster found in the comparison catalogue.

These definitions are applied independently for each external catalogue. An \astrolink\ cluster may therefore be reciprocal with respect to one catalogue, fragmented relative to another, and unmatched in a third.

\subsubsection{Assignment of structural type}
To assign a single structural label to each \astrolink\ cluster, we combine matches across catalogues. When multiple catalogues provide conflicting classifications, we select the match that maximises the lower-bound Jaccard index,
\begin{equation} \label{eq:max_lower_bound_Jaccard}
    J(A_i, C_j \mid U) \quad \text{with} \quad U = S_A.
\end{equation}
This arbitration rule is the same as the optimisation strategy used for the dominant cluster catalogues (Secs.~\ref{subsubsec:hunt2024}~\&~\ref{subsubsec:ucc}) and mitigates spurious high-purity matches that would arise from stream overlaps via the \texttt{galstreams} catalogue. In practice, the ranking induced by the lower-bound Jaccard index closely tracks that of the upper-bound Jaccard index for globular cluster and dwarf galaxy catalogues, so this choice has negligible impact on those systems. \astrolink\ clusters with no match to any catalogue are simply labelled as \textit{unknown}.

This fixed-$S$ bi-directional construction converts our availability-aware comparison scheme into a discrete structural taxonomy. Whereas Sec.~\ref{subsec:catalogues} assessed global matching performance as a function of $S$, the present analysis characterises the morphological and hierarchical relationships between the \astrolink\ clustering and external catalogues at the adopted operating point.

\begin{table*}[t] \label{tab:hierarchy_relationship_classification}
    \centering
    \resizebox{\textwidth}{!}{%
        \begin{tabular}{crrrrrrrrrr}
            \toprule
            Map direction & All & True 1-1 & Frag 1-1 & Merg 1-1 & F \& M 1-1 &
            True Frag & True Merg & Frag \& Merg & Isolated & No match \\
            \midrule
            \midrule
            
            \multirow{3}{*}{\makecell[c]{Hunt2024 \\ $\downarrow$ \\ Ours}}
             & 7167 & 4101 & 426 & 201 & 34 &   & 634 &   & 69 & 1702 \\
             & 4282 & 3742 & 361 & 153 & 25 &   &     &   & 1  &      \\
             & 2550 & 2307 & 198 & 36  & 9  &   &     &   &    &      \\
            \midrule
            
            \multirow{3}{*}{\makecell[c]{UCC \\ $\downarrow$ \\ Ours}}
             & 18299 & 2632 & 192 & 1839 & 166 &   & 6007 & 3 & 60 & 7400 \\
             & 5845  & 2491 & 177 & 1650 & 148 &   & 1375 & 2 & 2  &      \\
             & 3236  & 1431 & 80  & 1017 & 64  &   & 643  & 1 &    &      \\
            \midrule
            
            \multirow{3}{*}{\makecell[c]{Vasiliev2021 \\ $\downarrow$ \\ Ours}}
             & 170 & 55 & 88 & 1 &   &   & 3 &   & 1 & 22 \\
             & 124 & 41 & 82 & 1 &   &   &   &   &   &    \\
             & 92  & 28 & 63 & 1 &   &   &   &   &   &    \\
            \midrule
            
            \multirow{3}{*}{\makecell[c]{Battaglia2021 \\ $\downarrow$ \\ Ours}}
             & 74 & 11 & 12 &   & 1 &   & 1 &   & 1 & 48 \\
             & 13 & 6  & 6  &   & 1 &   &   &   &   &    \\
             & 8  & 2  & 5  &   & 1 &   &   &   &   &    \\
            \midrule
            
            \multirow{3}{*}{\makecell[c]{galstreams \\ $\downarrow$ \\ Ours}}
             & 141 & 18 & 46 & 2 & 7 & 5 & 17 & 2 & 28 & 16 \\
             &     &    &    &   &   &   &    &   &    &    \\
             &     &    &    &   &   &   &    &   &    &    \\
            \midrule
            
            \multirow{3}{*}{\makecell[c]{Ours \\ $\downarrow$ \\ All others}}
             & 7410 & 3417 & 1117 & 476 & 111 & 4 & 770 & 97 & 17 & 1405 \\
             & 5027 & 3249 & 1022 & 400 & 92  &   & 224 & 36 & 4  &      \\
             & 3355 & 2272 & 726  & 263 & 47  &   & 47  &    &    &      \\
            \bottomrule
        \end{tabular}%
    }
    \caption{Hierarchy relationship classification between literature catalogues and the structures identified in this work at $S=3.8$. Each block corresponds to a specific comparison, with the arrow indicating the direction of mapping: objects in the upper catalogue are classified according to how they are recovered by the lower catalogue. Columns list the number of objects in each hierarchical relationship class, defined using best-match Jaccard associations: reciprocal one-to-one matches (True 1--1), reciprocal matches involving fragmentation and/or merging (Frag 1--1, Merg 1--1, F \& M 1--1), non-reciprocal fragmentation or merging (True Frag, True Merg, Frag \& Merg), isolated matches (non-reciprocal but without clear fragmentation or merging), and objects with no counterpart (No match). The ``All'' column gives the total number of objects considered in each sub-sample. Within each block, the first row lists all objects in the parent catalogue, the second row lists those with upper-bound Jaccard index $J_{\mathrm{upper}} > 0.5$, and the third row lists those with lower-bound Jaccard index $J_{\mathrm{lower}} > 0.5$. In the final block (``Ours $\downarrow$ All others''), each \astrolink\ cluster is assigned a single best-matching catalogue object by maximising the lower-bound Jaccard index across all catalogues, ensuring a unique structural classification per cluster.}
\end{table*}

\subsubsection{Global structural statistics}
The structural classifications defined above can be summarised by counting the multiplicities of reciprocal, fragmented, merged, isolated, and unmatched systems. Table~\ref{tab:hierarchy_relationship_classification} reports these counts at the adopted threshold $S=3.8$. For each external catalogue we list the number of catalogue clusters in each structural relationship class, together with the subset exceeding a representative quality threshold ($J>0.5$) under both the upper and lower availability bounds. This catalogue-centric view quantifies how frequently \astrolink\ recovers, fragments, or merges known structures of each type.

These statistics reveal clear differences between structure types. Open cluster catalogues (Secs.~\ref{subsubsec:hunt2024} and \ref{subsubsec:ucc}) are dominated by reciprocal one-to-one matches at high Jaccard values, indicating that compact and high-contrast systems are typically recovered intact. The UCC nevertheless contains many true merged and unmatched clusters, likely reflecting the larger fraction of lower-quality or loosely bound groups in that catalogue. In contrast, globular clusters and dwarf galaxies show a larger fractional contribution from reciprocal fragmentation classes, consistent with the hierarchical decomposition of extended systems at the leaf level (Secs.~\ref{subsubsec:vasiliev2021} and \ref{subsubsec:battaglia2021}). Similarly, many dwarf galaxies from \citet{Battaglia2021} appear as true merged or unmatched systems, which likely reflects the more limited coverage of these objects in our GDR3 subsample. The \texttt{galstreams} catalogue shows a comparable tendency toward fragmented or merged classifications. This reflects both the low contrast of tidal debris in our data space and the inherently one-sided nature of the comparison (Sec.~\ref{subsubsec:galstreams}), which allows overlaps with other structure types occupying only part of the full feature space.

The final block of Table~\ref{tab:hierarchy_relationship_classification} reverses the perspective by considering how the \astrolink\ clusters themselves are matched by the union of all catalogues. After resolving cross-catalogue ambiguities using the maximum lower-bound Jaccard index (Eq.~\ref{eq:max_lower_bound_Jaccard}), each \astrolink\ cluster is assigned a single best-matching catalogue system (if one exists). Applying the same classification then yields a global \astrolink-centric breakdown of how the hierarchy maps onto previously known structure. Most clusters participating in high-quality matches ($J_{\mathrm{upper}}>0.5$) fall into the reciprocal one-to-one or reciprocal fragmentation classes. This behaviour is expected given that \astrolink\ produces a hierarchical cluster tree and that its leaf clusters generally provide reasonable matches to most structure types (Sec.~\ref{sec:catalogue_crossmatching}). Purely non-reciprocal mergers or fragmentations are comparatively rare among high-Jaccard matches, suggesting that strong overlaps typically correspond to genuine structural associations rather than incidental footprint intersections. The resulting taxonomy therefore reflects astrophysical morphology as much as algorithmic behaviour. We do however take a closer look at individual structure types identified by our model-agnostic catalogue comparison method in the following sub-subsection.

\subsubsection{Known structures and potential discoveries}

\begin{figure*}
    \centering
    \includegraphics[trim={3mm 5mm 3mm 3mm}, clip, width=\linewidth]{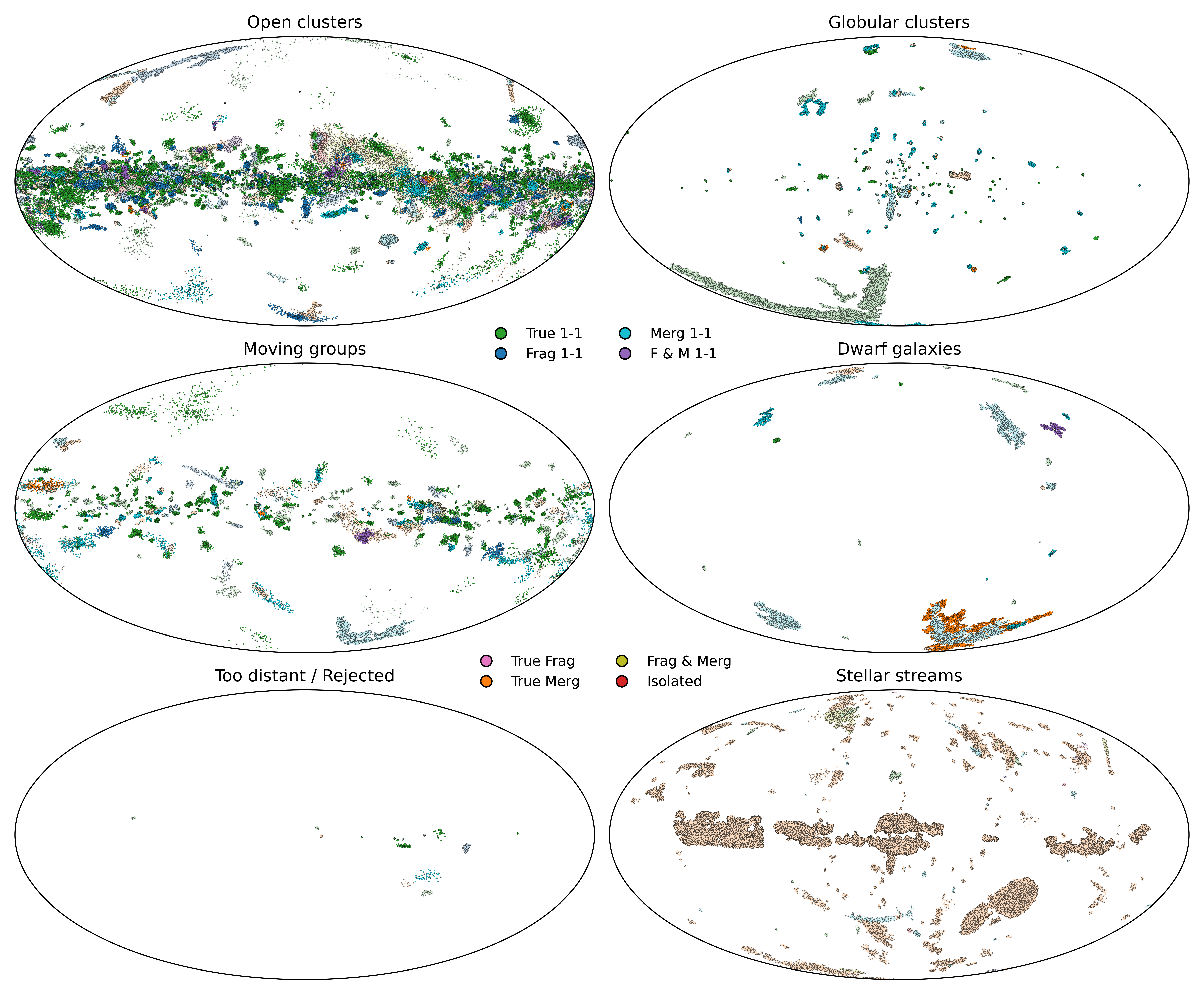}
    \caption{On-sky positions of stellar members of \astrolink\ clusters at $S=3.8$, shown in Galactic coordinates and separated by structure type as inferred from catalogue comparisons alone. Marker colours indicate the classification type (e.g. reciprocal matches, fragmentation, merging, or isolated), as defined in Sec.~\ref{subsubsec:structural_relationship_classes}. Markers are also whitened for clusters whose best match yields $J < 0.5$. Misclassifications are due to the comparison catalogues not containing a comprehensive representation of these structures -- which is often indicated by the poor structural mismatch between the comparison catalogues and our own (lighter, warmer colours).}
    \label{fig:structure_types_on_sky}
\end{figure*}

\begin{figure*}
    \centering
    \includegraphics[trim={12mm 3mm 12mm 3mm}, clip, width=\linewidth]{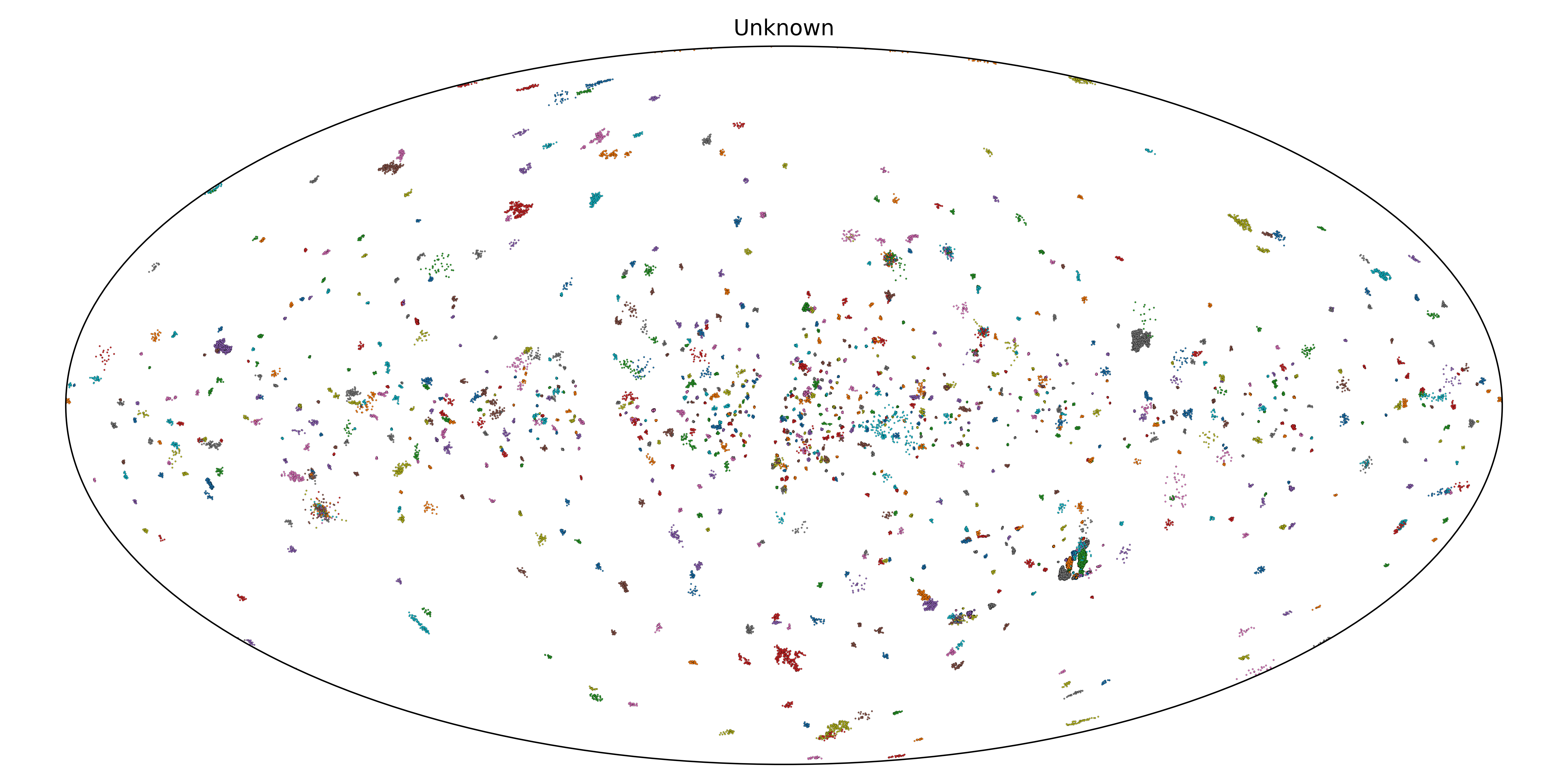}
    \caption{The same as Fig.~\ref{fig:structure_types_on_sky}, except here the 1414 \astrolink\ clusters with no overlap with any comparison catalogue are shown. Since there are no hierarchical relationships for unmatched clusters, marker colours are instead assigned from a ten colour cycle using the cluster label (as in Fig.~\ref{fig:all_clusters_on_sky}).}
    \label{fig:unknown_structures}
\end{figure*}

Figs.~\ref{fig:structure_types_on_sky}\&\ref{fig:unknown_structures} show the on-sky distribution of stellar members of \astrolink\ clusters at the adopted threshold of $S=3.8$, separated according to the structural and hierarchical classifications defined in Sec.~\ref{subsubsec:structural_relationship_classes}. In Fig.~\ref{fig:structure_types_on_sky}, clusters with identified counterparts in the comparison catalogues are grouped by their assigned relationship class, while Fig.~\ref{fig:unknown_structures} displays those clusters with no overlap with any catalogue. Together, these figures provide a spatial view of how the \astrolink\ hierarchy maps onto previously identified structures, as well as where it extends beyond them.

Clusters participating in reciprocal one-to-one matches are distributed broadly across the sky, with a clear concentration toward the Galactic plane where the majority of high-contrast stellar systems reside. Fragmented and merged systems are also present throughout this region, often co-located with areas of high stellar density. In contrast, isolated matches and lower-confidence associations appear more sparsely distributed, with no strong large-scale spatial coherence. The spatial overlap between different structural classes reflects the diversity of morphologies captured within the \astrolink\ hierarchy, ranging from compact systems to more extended or composite structures.

Across the different structure types represented in Fig.~\ref{fig:structure_types_on_sky}, clear differences are visible in both their spatial distributions and the strength of their crossmatches to the \astrolink\ hierarchy. Compact systems such as open and globular clusters are predominantly concentrated toward the Galactic plane and are most frequently associated with high-confidence matches, reflecting their relatively well-defined and localised structure in the data space. In contrast, dwarf galaxies and stellar streams occupy a broader range of Galactic latitudes and tend to exhibit more varied match strengths, with a larger fraction of fragmented or lower-confidence associations. Extended and diffuse structures are often represented by multiple \astrolink\ clusters or by clusters with reduced overlap under the adopted metrics, leading to a wider spread in their classification across the sky. And owing to the one-sided nature of the \texttt{galstreams} catalogue \citep{Mateu2023}, many apparent streams found by \astrolink\ are matched to other structural types. Conversely and for the same reasons, many \astrolink\ clusters identified as stellar streams have likely been done so falsely -- and are more likely to match to compact groups or composite structures that otherwise have no counterpart in the comparison catalogues\footnote{This is most notable with the galactic disc and Magellanic clouds that have been classified as stellar streams as well as with various stellar streams such as Sagittarius that have been classified as compact clusters. Such misclassifications occur because no catalogue we compare to contains a comprehensive representation of these structures. Of course this issue can be fixed by incorporating such a catalogue into our methodology. However, we refrain from doing so here since to our knowledge there are no such membership lists for these structures currently available and it is beyond the scope of this work to create them.}. The variation seen in Fig.~\ref{fig:structure_types_on_sky} highlights the differing degrees to which each structure type is captured by the \astrolink\ hierarchy at the adopted significance threshold, as well as the dependence of the crossmatching behaviour on intrinsic morphology and spatial extent.

The population of clusters with no catalogue counterpart, shown in Fig.~\ref{fig:unknown_structures}, spans a wide range of Galactic environments. These systems are not confined to any single region of the sky, appearing both along the Galactic plane and at higher latitudes. Their distribution suggests that the \astrolink\ clustering identifies a substantial number of structures that are either absent from, or not fully represented within, the comparison catalogues. These may include low-contrast, diffuse, or previously unreported systems.

\section{Conclusions and Outlook} \label{sec:conclusions}
In this work, we have applied the hierarchical, model-agnostic clustering algorithm \astrolink\ to a locally complete subsample of the GDR3 catalogue comprising over one billion stars with five-dimensional astrometric information. By incorporating a correction for the observational selection function, a distance contraction designed to stabilise uncertainties, and a physically motivated feature scaling, we constructed a data representation suited to extracting meaningful clustering structure from this large data set. At an adopted significance threshold of $S=3.8$, \astrolink\ identifies $7410$ statistically significant clusters spanning a wide range of sizes and morphologies, thereby providing a global, multi-scale view of the Milky Way’s stellar structure.

Through comparison with a range of external catalogues, including open clusters, globular clusters, dwarf galaxies, and stellar streams, we find that \astrolink\ recovers known structures with a high degree of fidelity while naturally capturing their diversity in morphology. Compact and high-contrast systems, such as open and globular clusters, are frequently identified as reciprocal one-to-one matches, whereas more extended systems, including dwarf galaxies and stellar streams, are often represented by multiple clusters or by clusters that capture only their densest regions. This behaviour is consistent with the hierarchical nature of the \astrolink\ output and reflects the intrinsic structural complexity of these systems rather than a limitation of the method.

The structural taxonomy introduced in this work provides a unified and model-independent framework for interpreting the relationships between clusters identified by \astrolink\ and those reported in the literature. By constructing a bi-directional best-match relation and classifying associations into reciprocal, fragmented, merged, isolated, and unmatched categories, we are able to characterise not only the degree of agreement between catalogues, but also the ways in which different structure types are represented within a hierarchical clustering paradigm. This approach highlights that many astrophysical systems are not well described as single-scale objects, but instead exhibit internal structure that is resolved at different levels of the hierarchy.

A substantial population of \astrolink\ clusters is found to have no counterpart in the comparison catalogues. These systems are distributed across a wide range of Galactic environments and may correspond to low-contrast or diffuse structures, substructure within previously identified systems, or objects that have not yet been catalogued. While the present analysis does not establish the physical nature of these clusters, it demonstrates the capacity of a model-agnostic clustering approach to extend beyond existing catalogues and to identify candidate structures for further investigation.

Several limitations of this study should be noted. The clustering is performed using a five-dimensional astrometric space without radial velocities, and therefore does not fully capture the dynamical structure of the Galaxy. In addition, the results depend on the adopted data transformations, including the distance contraction and feature scaling, as well as on the estimation of the selection function. The comparison to external catalogues is also inherently limited by their completeness and by differences in how membership is defined, particularly in the case of diffuse structures such as stellar streams.

Future work will focus on extending this framework to higher-dimensional data by incorporating radial velocity measurements. Additionally, the clustering results of the current 5D and upcoming 6D applications are intended to be improved with the propagation of feature space uncertainties into the clusters themselves. These results will lead us to explore the hierarchical structure in greater detail. In particular, a systematic analysis of the unmatched clusters and a deeper investigation of the physical interpretation of fragmentation and merging within the \astrolink\ hierarchy will provide further insight into the assembly history and structure of the Milky Way.

\section*{Data availability}

The code used to produce the results of this work is available on the \astrolink\ GitHub at \url{https://github.com/william-h-oliver/astrolink}. The \astrolink\ stellar and cluster catalogues produced in this work are available from Zenodo at \url{https://doi.org/10.5281/zenodo.23114479}.

\begin{acknowledgements}
      This work is funded by the Carl-Zeiss-Stiftung through the NEXUS program. This work was supported by the Deutsche Forschungsgemeinschaft (DFG, German Research Foundation) – 573580088.
\end{acknowledgements}

%
%

\bibliographystyle{aa}
\bibliography{references}

\end{document}